\documentclass{article}
\RequirePackage{amsthm,amsmath,amsfonts,amssymb}
\RequirePackage[authoryear]{natbib}
\RequirePackage[colorlinks,citecolor=blue,urlcolor=blue]{hyperref}
\RequirePackage{graphicx}
\usepackage{xcolor}
\usepackage{comment}
\usepackage{booktabs}
\usepackage{longtable}
\usepackage{array}
\usepackage{bbm}
\usepackage{algorithm}
\usepackage{algpseudocode}
\usepackage{enumerate}
\usepackage[a4paper, total={6in, 8.5in}]{geometry}

\theoremstyle{plain}

\theoremstyle{remark}

\usepackage{amssymb} 

\title{Topic-informed dynamic mixture model
for occupational heterogeneity in health risk behaviors}

\author{Lorenzo Schiavon, \and Mattia Stival, \and Angela Andreella, \and Stefano Campostrini \\
{\small Department of Economics,
Ca' Foscari University of Venice, Venice, Italy }}

\date{}

\begin{document}

\maketitle

\begin{abstract}
Behavioral risk factors, i.e., smoking, poor nutrition, alcohol misuse, and physical inactivity (SNAP), are leading contributors to chronic diseases and healthcare costs worldwide. Their prevalence is shaped 
by demographic characteristics 
and also by contextual ones such as socioeconomic and occupational environments.
In this study, we leverage data from the Italian health and behavioral surveillance system PASSI to model SNAP behaviors through a Bayesian framework that integrates textual information on occupations. We use Structural Topic Modeling (STM) to cluster free-text job descriptions into latent occupational groups, which inform mixture weights in a multivariate ordered probit model. 
Covariate effects are allowed to vary across occupational clusters and evolve over time. To enhance interpretability and variable selection, we impose non-local spike-and-slab priors on regression coefficients. Finally, an online learning algorithm based on sequential Monte Carlo enables efficient updating as new data become available.  This dynamic, scalable, and interpretable approach permits observing how occupational contexts modulate the impact of socio-demographic factors on health behaviors, providing valuable insights for targeted public health interventions.\\

\end{abstract}


\section{Introduction}
Behavioral risk factors such as smoking, poor nutrition, alcohol misuse, and physical inactivity, collectively referred to as SNAP, are major contributors to chronic diseases and global mortality \citep{who2014global}. These behaviors not only compromise individual health but also significantly burden healthcare systems \citep{tarricone2006}. In Italy, \citet{possenti2024} estimated that smoking was responsible for one-third of hospitalizations among adults over 30 in 2018, costing \texteuro 1.64 billion, while obesity-related costs reached \texteuro 13.34 billion in 2020 \citep{derrico2022}. In a review of 26 studies, \citet{manthey2021} estimated that alcohol use alone accounts for approximately 1.5\% of GDP.
Understanding how risk factors emerge and coalesce is key to designing effective, targeted interventions to impact public health outcomes \citep{prochaska2011review}. 

SNAP behaviors are known to correlate with demographic variables such as age, sex, and geographical location \citep{flaskerud2012social, minardi2011social}. 
Thus, modeling individual propensity to SNAP through appropriate regression techniques helps explain population variability in SNAP levels, collected in the matrix $\boldsymbol{Y}$, based on demographic covariates $\boldsymbol{X}$ \citep{kuntsche2017binge,paavola2004smoking}.

Beyond macro-environmental influences, risk behaviors are often most affected by the socioeconomic micro-environment in which an individual lives, such that we expect a heterogeneous effect $\boldsymbol{B}$ of the demographic characteristic $\boldsymbol{X}$ across different environments.
In particular, the occupational environment, along with socioeconomic status, plays a key role in shaping a person’s social and economic context. In adulthood, occupation not only determines time use and exposure to stressors, but also represents a primary source of social interactions and lifestyle norms, thereby shaping individual habits \citep{clougherty2010work, marmot2020social}.

We analyzed the occurrence of SNAP behaviors in PASSI data \citep{baldissera2011peer}, a monthly-based sample survey that gathers information on behavioral risk factors, socio-demographic characteristics, and self-reported non-communicable diseases in Italy.
Using these data to understand the link between occupation and behavioral risk factors may inform targeted public health policies. While job strain is often studied as a driver of unhealthy behaviors \citep{nyberg2013job}, broader associations between SNAP factors and job type remain understudied. 
Existing research typically focuses on specific professions \citep{ficarra2011tobacco}, 
behaviors like alcohol or smoking \citep{terza2002alcohol, mullahy1996employment, 
chiatti2010cigarette}, employment status \citep{schneider2011impact, minelli2014employment}, or age groups \citep{
bonanomi2022employment}.
However, an integrated understanding of how multiple risk factor behaviors depend on both socioeconomic and occupational context, demographic drivers, and temporal trends is still lacking, often due to data limitations.
These are overcome in this study since we use the PASSI surveillance system, which offers rich information on both socio-demographic characteristics and lifestyle habits of each respondent. 
In addition, PASSI collects occupational data through both an open-ended employment question and the ISTAT occupation classification and sectors.
We assume that integrating occupational information together with socioeconomic traits can provide a more comprehensive representation of an individual's social micro-environment. 

For each respondent $n$, we define a vector of words $\boldsymbol{w}_n$ obtained by combining the open-ended employment response with the textual descriptions from the ISTAT classification and occupational sectors. 
We also define a vector $\boldsymbol{v}_n$ containing socioeconomic indicators.
To enable a quantitative, data-driven characterization of occupations while preserving the richness of this information, we employ topic modeling techniques that incorporate exogenous covariates. 
These methods summarize the textual and socioeconomic information into homogeneous clusters of individuals, with $s_n$ indicating the cluster membership of respondent $n$.
We expect these clusters to reflect both the job description $\boldsymbol{W}$ and groups of individuals characterized by different relationships $\boldsymbol{B}$ between demographic factors and SNAP, as illustrated in the model scheme reported in Figure \ref{fig:model-dag}. 

\begin{figure}[ht]
    \centering
    \includegraphics[width=.7\linewidth]{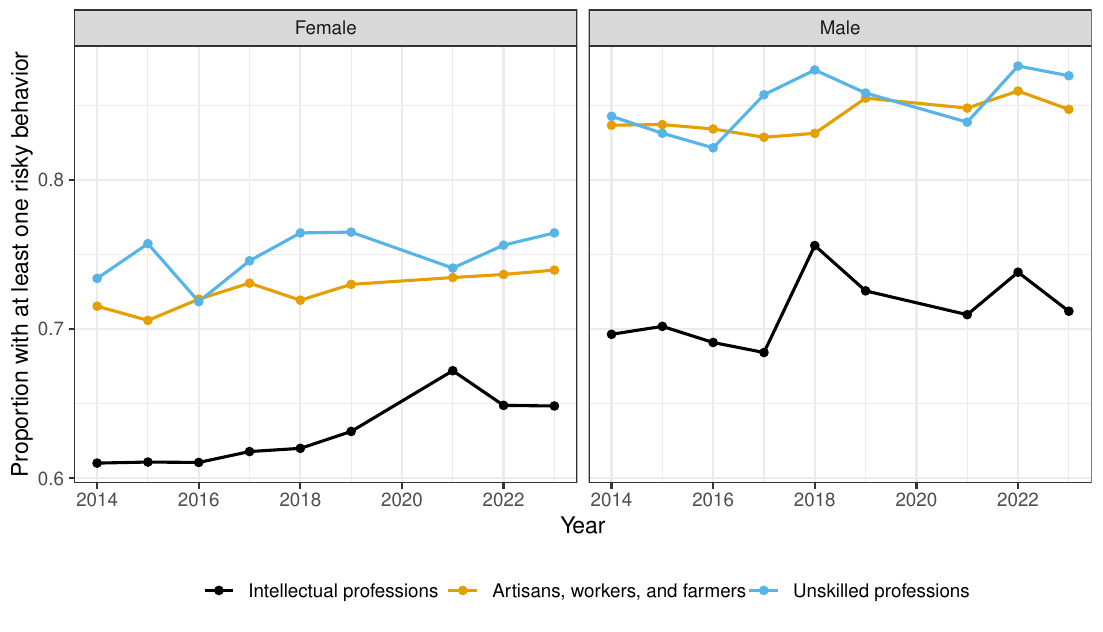}
    \caption{Proportions of individuals with at least one risky behavior are reported by sex, occupation, and year.}
    \label{fig:ts_eda}
\end{figure}

Following societal shifts, the influence of social and demographic factors on risky health behaviors evolves over time \citep{teng2020}. For example, although such behaviors were traditionally more prevalent among men, decreasing gender inequalities and shifting social norms have led to more balanced patterns across sexes \citep{white2020}.
These evolving associations highlight the need for a time-varying coefficient matrix $\boldsymbol{B}^{(t)}$ where $t = 1, \ldots, T$.
These effects are expected to vary across occupational and socioeconomic groups, reflecting differences in education and changes in work environments.
Technological automation has increased sedentary behavior in manual labor jobs \citep{kumareswaran2023}, while the COVID-19 pandemic accelerated the adoption of remote work, particularly in higher-educated and higher-skilled sectors. This latter shift has altered daily routines and health behaviors such as diet and physical activity, with uneven impacts across socioeconomic and occupational groups \citep{lyzwinski2024}. 
Figure~\ref{fig:ts_eda} illustrates short-term variation in risk trajectories for three job categories among those defined by \cite{istat_cp2011}.
These trends motivate a model in which regression coefficients vary over time and across latent micro-environment groups, and where parameters can be updated continuously as new data become available.

We summarize our modeling goals as follows. A suitable statistical framework should: 
(1) model the relationship between demographic variables $\boldsymbol{x}_n$ and SNAP outcomes $\boldsymbol{y}_n$ as varying across time $t$ and latent groups $s_n \in \{1,\ldots, K\}$ to capture heterogeneity; 
(2) incorporate textual occupational descriptions $\boldsymbol{w}_n$ and socioeconomic indicators $\boldsymbol{v}_n$ through structured topic modeling to inform group membership; 
(3) support online parameter updates as new data become available; and 
(4) enforce sparsity in the effects, since not all covariates are expected to influence SNAP outcomes in every group or time period.
To meet these goals, we propose a dynamic mixture of multivariate ordered probit models, where the latent outcome is centered at $\boldsymbol{x}_n \boldsymbol{B}_k^{(t)}$, with $\boldsymbol{B}_k^{(t)}$ denoting the group- and time-specific coefficient matrix. 
The proposed framework aims to offer a novel perspective by explicitly embracing the inherent complexity of the data under analysis \citep{rod2023complexity}.

The manuscript is organized as follows. Section~\ref{model} introduces the model,
describing how textual occupational and socioeconomic information inform the mixture weights using the Structural Topic Model \citep[STM,][]{roberts2016}. 
Sparsity and temporal dynamics in $\boldsymbol{B}_k^{(t)}$ are modeled via a spike-and-slab prior with a non-local slab formulation \citep{johnson2010}. 
Section~\ref{sec:algo} outlines a Sequential Monte Carlo (SMC)-based online learning algorithm, assessed through simulation studies in Section~\ref{simulation}. 
Section~\ref{application} applies the model to data from the Italian health surveillance system PASSI \citep{baldissera2011peer}, followed by discussion and conclusions in Section~\ref{discussion}.

\section{Informed dynamic mixture of multivariate ordered probit regression}
\label{model}

\subsection{Model specification}\label{subsec:ms}
In this subsection, we present the general framework of a dynamic mixture of multivariate ordered probit models for health-related behaviors. The incorporation of occupational and socioeconomic information through a Structural Topic Model, as well as the dynamic shrinkage prior for the regression coefficients, are described in the subsequent subsections.

Let $\boldsymbol{y}_n = [y_{n1}, \dots, y_{nP}]$ denote the response vector for respondent $n$, where the $p$-th entry $y_{np}=c+1$ indicates the ordinal level $c$ of the risky behavior $p \in \{1, \dots, P\}$ for respondent $n \in \{1, \dots, N\}$, with $c \in \{0, \ldots, C_p - 1\}$ and $C_p$ denoting the number of distinct categories for risky behavior $p$. 
Without loss of generality, the marginal ordinal variable $y_{np}$ can always be represented  as 
$$
y_{np} =  \sum_{c=0}^{C_{p-1}} \mathbbm{1}(z_{np} >\tau_{pc} ),
$$
in terms of a latent variable $z_{np} \sim F$ where $F$ is a suitable distribution, and threshold $\tau_{pc}$ is defined such that $\tau_{p(c-1)}<\tau_{pc} \in \mathbb{R}$ for $c \in \{1, \ldots, C_p - 1\}$ and $\tau_{p0} = -\infty$, $\tau_{pC_p} = \infty$ for any $p=1,\ldots,P$. 
Common models such as ordered probit \citep{aitchison1957} and logistic \citep{mccullagh1980} regressions exploit this representation by assuming $F$ belongs to the Gaussian or logistic distribution families, respectively.

For each respondent, we are interested in exploring the relationship between the SNAP level $\boldsymbol{y}_n$ and their demographic profile described by the covariate vector $\boldsymbol{x}_n = [x_{n1}, \dots, x_{nQ}]$, including characteristics as age class, sex, and the Italian macro-region of residence. In this notation, $x_{nq}$ denotes the $q$-th covariate for subject $n$. As previously introduced, we expect this relationship to vary across the population. 

To account for such heterogeneity, we model the vector $\boldsymbol{y}_n$ using a mixture of $K$ multivariate ordered probit regression models
\begin{equation}
\label{eq:model}
    {y}_{np} =  \sum_{c=0}^{C_{p-1}} \mathbbm{1}(z_{np} >\tau_{pc} ),
    \quad
    \left(\boldsymbol{z}_{n} \vert
    \boldsymbol{x}_n, s_n = k,    \boldsymbol{B}_{k }^{(t)}\right) \sim \mathcal{N}_P(\boldsymbol{x}_n \boldsymbol{B}_{k}^{(t)}, \boldsymbol{I}_P),
\end{equation}
where $\boldsymbol{z}_n = [z_{n1}, \ldots, z_{nP}]$ is a $P$-variate latent Gaussian vector, conditional on $s_n$, which identifies the latent mixture component for respondent $n$.
$\boldsymbol{x}_n \boldsymbol{B}_k^{(t)}$ is then the mean of $\boldsymbol{z}_n$, given the respondent's group allocation $s_n = k$ where $k \in \{1, \ldots, K\}$ and its associated matrix of regression coefficients $\boldsymbol{B}_k^{(t)} \in \mathbb{R}^{Q \times P}$.
Notably, the matrix of coefficients is indexed by ${(t)}$, indicating the time instant (month) in which data are collected, reflecting dynamic patterns within and between the years. For simplicity in notation, this time indexing is not reported on the quantities indexed by the respondent's index $n$, implicitly referring with $n$ also to the time $(t_n)$ in which the data were collected.

\begin{figure}[ht]
    \centering
    \includegraphics[width=0.5\linewidth]{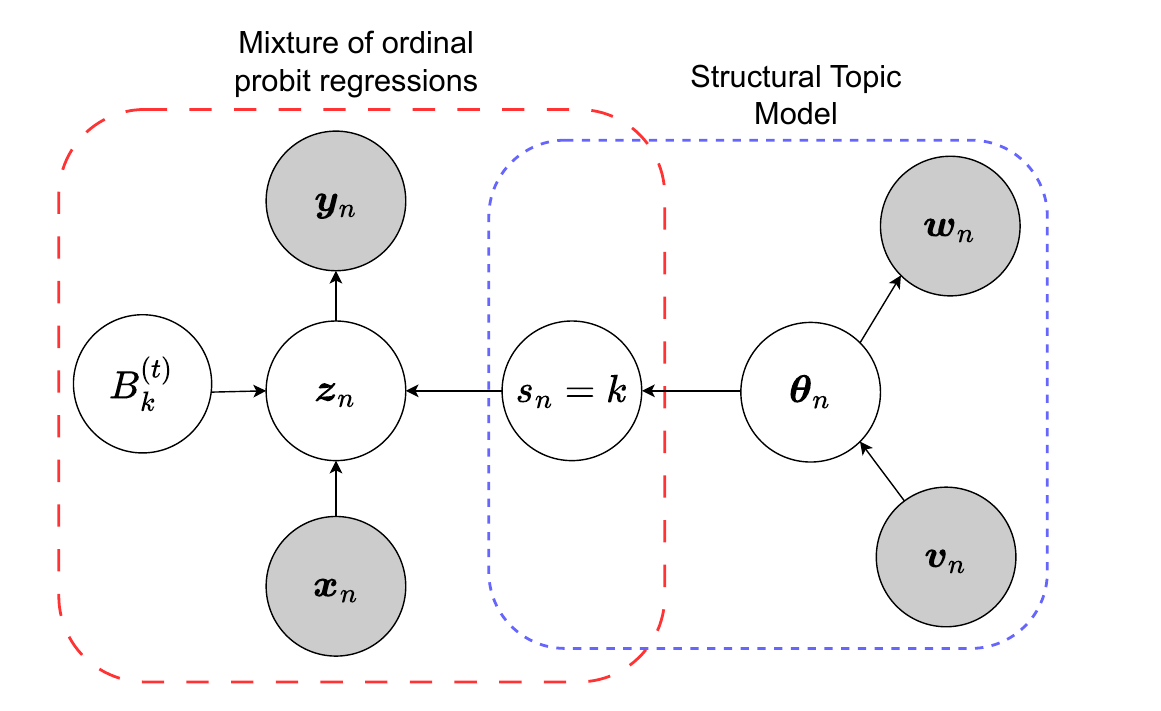}
    \caption{Model diagram describing the role of $s_n$: a bridge between the topic model and the mixture of probit regression models. Observed data vectors are colored in grey.}
    \label{fig:model-dag}
\end{figure}

The mixture formulation explores whether the propensity for risky behaviors, conditional on the demographic characteristics, varies across the population.
We expect this heterogeneity to be driven by the context in which individuals live. 
Thus, if additional information is available, this can be used to model the prior knowledge about the group membership of the individuals. 
In our framework, both occupational text \(\boldsymbol{w}_n\) and socioeconomic characteristics \(\boldsymbol{v}_n\) jointly inform group membership \(s_n\) through the STM, defining a unified socioeconomic construct that captures occupational micro-environments. Socioeconomic indicators are incorporated at this stage rather than directly in the SNAP regressions, thereby avoiding redundancy.
Figure~\ref{fig:model-dag} illustrates the relationships among all the components described above, highlighting how the respondent’s group allocation $s_n = k$ links the topic model to the mixture of probit regression models.

\subsection{Topic modeling informed prior for group membership}
\label{subsec:tm}

The occupational environment, often reflecting an individual's broader socioeconomic context, plays a key role in shaping health behaviors \citep{clougherty2010work,  marmot2020social}.
Unlike other contextual factors, such as family or community setting, occupational information is routinely collected in health surveillance systems like PASSI. 

In light of this, we hypothesize the existence of latent clusters representing distinct socioeconomic and occupational groups, which influence individuals’ propensity to engage in risky behaviors. We then model the prior distribution of the mixture weights, $\Pr(s_n = k)$ with $k \in \{1, \ldots, K\}$, by incorporating occupational information. 
We assume that respondents with similar occupations, particularly in socioeconomic terms, are more likely to be assigned to the same mixture component.
However, open-ended responses may be incomplete or vague, while relying solely on coded classifications could produce an excessive number of small, non-representative categories. 
Thus, we link the textual description of the occupation of one respondent to its latent representation $s_n$, which defines, in our setting, both the occupational group and the \emph{a priori} behavioral group. To enhance the grouping of textual information, we also leverage individual socioeconomic indicators, i.e., education level and economic situation, stored in the $N \times U$ matrix $\boldsymbol{V}$.

To model the combined textual and socioeconomic data, we employ the STM proposed by \citet{roberts2016}, which allows the inclusion of exogenous covariates. 
Consistent with the authors, we specify
\begin{equation}\label{eq:tm1}
    s_n \vert \boldsymbol{\theta}_n \sim \text{Multinomial}_K(1, \boldsymbol{\theta}_n), \quad \boldsymbol{\theta}_n \sim \text{LogisticNormal}_{K-1}( \boldsymbol{\Lambda}^\top \boldsymbol{v}_n , \boldsymbol{\Psi}),
\end{equation}
where $\boldsymbol{\theta}_n$ is a $K$-dimensional vector defined on the simplex $(0,1)^K$. 
Here, $\boldsymbol{v}_n$ denotes the $n$-th row of the covariate matrix $\boldsymbol{V}$, while $\boldsymbol{\Lambda}$ is a $(K-1) \times U$ coefficient matrix and $\boldsymbol{\Psi}$ is a positive semi-definite $(K-1)\times(K-1)$ covariance matrix, both treated as hyperparameters.

For the textual information, the data-generating process is then completed by defining a suitable distribution over the $H$ words in the vocabulary of the textual collection. The $i$-th word $w_{ni}$ in the text of subject $n$ is sampled as:
\begin{equation}\label{eq:tm2}
    w_{ni} \vert  \zeta_{ni} \sim \text{Multinomial}_H(1, \boldsymbol{\gamma}_{\zeta_{ni}}), \quad \zeta_{ni} \vert \boldsymbol{\theta}_n \sim \text{Multinomial}_K(1, \boldsymbol{\theta}_n),
\end{equation}
where $\boldsymbol{\gamma}_{k}$
are cluster-specific hyperparameter vectors set on the simplex $(0,1)^{H}$.

While we refer to
\citet{roberts2016} for other modeling details, it is relevant to highlight the dual role of the topic proportion vector $\boldsymbol{\theta}_n$. 
On one hand, it defines the probability distribution of topic $\zeta_{nh}$ for each word within the text of respondent $n$; then, conditional on $\zeta_{ni}$, a word is selected from the corresponding distribution, i.e., Equation \eqref{eq:tm1}. On the other hand, being $\boldsymbol{\theta}_n$ also the probability of the latent occupational group $s_n$, it also represents the \emph{a priori} behavioral group, i.e., Equation \eqref{eq:tm2}. 
Suppose no data on SNAP behaviors are observed. In that case, the posterior inference on $\boldsymbol{\theta}_n$ given the texts can be achieved by considering the STM implementation 
in the {\sf R} package \texttt{stm} \citep{stm-r}.
By exploiting variational techniques, approximated posterior inference can be quickly conducted even with large datasets.

Notably, the cluster allocation $s_n$ is the only term of STM that is involved in the mixture of probit regression models. 
As later explained in Section \ref{sec:algo}, this enables us to perform two-stage modular inference by sequentially estimating the STM and adjusting the estimates based on behavioral data.

\subsection{Dynamic nonlocal prior for regression coefficients}
\label{subsec:beta-prior}
When considering the regression parameters, we expect two main features. 
First, certain risky behaviors within occupational groups are likely to evolve over time. 
Second, among all regression parameters, only a subset is expected to have a meaningful impact, while many effects are negligible. 
To capture both aspects, we specify a dynamic prior that induces sparsity in the regression coefficients.

To model temporal dependence, we define a time-dynamic process for each entry $\beta_{kpq}^{(t)}$ of $\boldsymbol{B}_k^{(t)}$, representing the effect at time $t$ of the $q$-th covariate in $\boldsymbol{X}$ on outcome $p$ for group $k$. We assume a Markovian dependence structure, i.e., 
 $p(\beta_{kpq}^{(t)} \vert  \beta_{kpq}^{(1)}, \ldots, \beta_{kpq}^{(t-1)}) = p(\beta_{kpq}^{(t)} \vert  \beta_{kpq}^{(t-1)})$.

To induce sparsity, we place spike-and-slab type priors on the elements $\beta_{kpq}^{(t)}$, which shrink coefficients with weak or uncertain effects toward zero.
To enhance mixing and improve parameter identification during estimation, we follow \citet{ishwaran2005} and adopt a continuous shrinkage prior with density 
\begin{equation}\label{eq:shr_prior}
    p(\beta_{kpq}^{(t)} \vert \beta_{kpq}^{(t-1)}) = \pi_0(\beta_{kpq}^{(t-1)}) f_{\mathcal{N}}(\beta_{kpq}^{(t)}; 0, \sigma_0^2) + \{1-\pi_0(\beta_{kpq}^{(t-1)})\} f_{\textnormal{slab}}(\beta_{kpq}^{(t)}),
\end{equation}
where $\pi_0(\beta_{kpq}^{(t-1)}): \mathbb{R}\rightarrow (0,1)$ is a function of the past parameter. 
The spike density $f_{\mathcal{N}}(\beta_{kpq}^{(t)}; 0, \sigma_0^2)$ is assumed to be a Gaussian density centered at zero with small variance $\sigma_0^2$, ensuring substantial prior mass is placed near zero. 
In contrast, the slab distribution $f_{\textnormal{slab}}(\beta_{kpq}^{(t)})$ is specified as a nonlocal distribution with respect to $\beta_0 = 0$.

The nonlocal slab assigns negligible prior mass near zero \citep{johnson2010} and concentrates probability away from the origin on both the positive and negative axes. This structure sharpens the distinction between signal and noise while simultaneously separating negative from positive effects. As a result, regression coefficients naturally fall into three interpretable macro-categories: negative, negligible, and positive. This facilitates interpretation and the identification of population clusters with distinct patterns of covariate influence on risky behaviors.
%
Operatively, we follow \citet{denti2023} and specify $f_\textnormal{slab}(\beta_{kpq}^{(t)})$ as a bimodal mixture of two Gaussians $\mathcal{N}(\mu_{-1}, \sigma_{-1})$ and $\mathcal{N}(\mu_{1}, \sigma_{1})$, 
and suitably weighted by a function 
that reduces mass near the origin.
This induces a three-component mixture prior 
\begin{equation}
\label{eq:beta-prior}
    p(\beta_{kpq}^{(t)} \vert \beta_{kpq}^{(t-1)}) = 
\pi_{0}(\beta_{kpq}^{(t-1)})  f_{\mathcal{N}}(\beta_{kpq}^{(t)}; 0, \sigma_0^2) + \sum_{l \in \{-1,1\}} \pi_{l}(\beta_{kpq}^{(t-1)})
    g_{l}(\beta_{kpq}^{(t)}; \mu_{l}, \sigma_{l}, \xi), 
\end{equation}
with $\pi_l(\beta_{kpq}^{(t-1)}) \in (0,1)$ 
for $l \in \{-1,0,1\}$ and such that $\sum_{l=-1}^{1} \pi_l(\beta_{kpq}^{(t-1)}) = 1$. 
The functions $g_{-1}(\beta_{kpq}^{(t)}; \mu_{-1}, \sigma_{-1}, \xi)$ and $g_{1}(\beta_{kpq}^{(t)}; \mu_{1}, \sigma_{1}, \xi)$ are kernel functions allocating higher mass of probability on the negative and positive semi-axes, respectively. 
The detailed prior specification 
is reported in the SM, Appendix B.

To model coefficient dynamics, we link the prior mixture probabilities $\pi_{l}(\beta_{kpq}^{(t-1)})$ at time $t>1$ to the coefficient value $\beta_{kpq}^{(t-1)}$ observed at time $t-1$. 
We specify increased probability that each coefficient at time $t$ of remaining in the same mode of time $t-1$, inducing smooth temporal dynamics and avoiding abrupt changes, 
as the dynamic spike-and-slab processes \citep{rockova2021}.
The precise specification of the mixture probabilities and their evolution is provided in Appendix B of the SM, together with illustrative examples of the resulting prior coefficient densities (Figure S2).
For lower-dimensional parameterized models, Gaussian random walk priors on $\beta_{kpq}^{(t)}$ are also an option worth being considered.

\section{Forward estimation via Sequential Monte Carlo}
\label{sec:algo}

\subsection{Modular framework for sequential inference}\label{sec:algo3.1}

Given the sequential nature of the health surveillance systems' data collection, our goal is to infer the joint distribution of group indicators and cluster-specific regression parameters as new data become available. We then first introduce key notation and emphasize the modular structure of the model. We then develop a forward inferential strategy to obtain our target inferences efficiently.

For a time of interest $T$, we denote with $\boldsymbol{B}^{(T)}$ the set of group coefficient matrices $\boldsymbol{B}^{(T)} = \{\boldsymbol{B}_1^{(T)},\ldots,\boldsymbol{B}_K^{(T)}\}$ and with $\boldsymbol{s}^{(T)}$ the membership indicators $\boldsymbol{s}^{(T)}=(s_1^{(T)},\ldots,s_{N_T}^{(T)})$.
For all the elements considered, the superscript $(1:T)$ will indicate all time points $t \in \{1, \dots, T\}$.
For instance, $\mathcal{D}^{(1:T)}$ will refer to all the observed data up to time $T$, i.e., $\mathcal{D}^{(1:T)} = \{\boldsymbol{X}^{(1:T)},\boldsymbol{V}^{(1:T)}, \boldsymbol{Y}^{(1:T)}, \boldsymbol{W}^{(1:T)}\}$.
We remark that, in 
our study, multiple observations are gathered at each time point $T$. Specifically, $\mathcal{D}^{(T)} =  \{\boldsymbol{X}^{(T)}, \boldsymbol{V}^{(T)}, \boldsymbol{Y}^{(T)}, \boldsymbol{W}^{(T)}\}$ denotes the collection of data corresponding  to the $N_{T}$ respondents whose information becomes available at time $T$. In the case of the PASSI survey, where data are collected on a monthly basis, $N_t$ is typically on the order of several thousand observations.

Our goal is to recover the \emph{filtering} distribution  
$p(\boldsymbol{s}^{(T)}, \boldsymbol{B}^{(T)} \vert \mathcal{D}^{(1:T)})$, for any $T$.
As is common in Bayesian inference, we approximate the target distribution via a Monte Carlo method. Specifically, we design an algorithm capable of sampling a sufficiently large number of particles from $p(\boldsymbol{s}^{(1:T)}, \boldsymbol{B}^{(1:T)} \vert \mathcal{D}^{(1:T)})$, and then numerically integrate out the non-relevant parameters $s^{(1:T-1)}$ and $\boldsymbol{B}^{(1:T-1)}$.

Conditionally on the covariates $\boldsymbol{X}^{(1:T)}$, $\boldsymbol{V}^{(1:T)}$, and on the group memberships $\boldsymbol{s}^{(1:T)}$, the model described in Section~\ref{model} assumes two independent data-generating processes: one for the levels of risky behaviors, $\boldsymbol{Y}^{(1:T)}$, and another for the occupational text descriptions, $\boldsymbol{W}^{(1:T)}$.
This assumption leads to the following decomposition
\begin{align}
\label{eq:post-dec}
    p(\boldsymbol{s}^{(1:T)}, \boldsymbol{B}^{(1:T)} \vert \mathcal{D}^{(1:T)}) = &p(\boldsymbol{Y}^{(1:T)} \vert \boldsymbol{B}^{(1:T)}, \boldsymbol{s}^{(1:T)}, \boldsymbol{X}^{(1:T)}) p(\boldsymbol{Y}^{(1:T)}, \boldsymbol{W}^{(1:T)})^{-1}\\
    &p(\boldsymbol{B}^{(1:T)}) \,
    p(\boldsymbol{s}^{(1:T)} \vert \boldsymbol{V}^{(1:T)},\boldsymbol{W}^{(1:T)}). \nonumber
\end{align}

Building on this factorization, estimation proceeds in two stages.  
In the first stage, we estimate the posterior distribution of group membership probabilities, given the textual and socioeconomic data, i.e.,
\begin{equation*}
\label{eq:prior}    
p(\boldsymbol{s}^{(1:T)} \vert \boldsymbol{W}^{(1:T)}, \boldsymbol{V}^{(1:T)} ) \propto \int_{\theta}
p( \boldsymbol{W}^{(1:t)} \vert \boldsymbol{\theta}^{(1:T)} 
)\, p(\boldsymbol{s}^{(1:T)} \vert \boldsymbol{\theta}^{(1:T)}) p(\boldsymbol{\theta}^{(1:T)} \vert \boldsymbol{V}^{(1:T)})  \, \textnormal{d} \theta.
\end{equation*}
In the second stage, inference is performed on the regression coefficients $\boldsymbol{B}^{(1:T)}$, while updating the latent group assignments $\boldsymbol{s}^{(1:T)}$ using the activity data $\boldsymbol{Y}^{(1:T)}$. 
The distribution $p(\boldsymbol{s}^{(1:T)} \mid \boldsymbol{W}^{(1:T)}, \boldsymbol{V}^{(1:T)})$ estimated in the first stage serves as an informative prior for the mixture weights in the multivariate ordinal regression model $p(\boldsymbol{Y}^{(1:T)} \mid \boldsymbol{B}^{(1:T)}, \boldsymbol{s}^{(1:T)}, \boldsymbol{X}^{(1:T)})$.

Efficient algorithms for estimating the STM provide accurate approximations of 
$p(\boldsymbol{s}^{(1:T)} \mid \boldsymbol{W}^{(1:T)}, \boldsymbol{V}^{(1:T)})$. 
Following \citet{roberts2016}, we employ computationally efficient logistic-normal approximations to the posterior topic proportions 
$p(\boldsymbol{\theta}^{(1:T)} \mid \boldsymbol{W}^{(1:T)}, \boldsymbol{V}^{(1:T)})$, 
which make sampling from $p(\boldsymbol{s}^{(1:T)} \mid \boldsymbol{W}^{(1:T)}, \boldsymbol{V}^{(1:T)})$ straightforward. 
This sampling depends 
on the hyperparameters 
$\boldsymbol{\Gamma}$, $\boldsymbol{\Lambda}$, and $\boldsymbol{\Psi}$, whose specification is discussed in detail in Subsection \ref{subsec:prior}. 
Finally, the joint posterior distribution $p(\boldsymbol{s}^{(1:T)}, \boldsymbol{B}^{(1:T)} \mid \mathcal{D}^{(1:T)})$ 
can be sequentially updated over time, as described in Subsection~\ref{sec:algo3.2}. 
Posterior samples can be obtained through Hamiltonian Monte Carlo (HMC)
\citep[e.g., via {\sf Stan},][]{carpenter2017stan}, Gibbs sampling \citep{gelfand1990}, 
or SMC methods \citep{kantas2015particle}. 
In Subsection~\ref{subsec:seq-mc}, we detail the custom SMC procedure developed for this study, 
which achieves substantial computational efficiency compared to alternative approaches, as shown in the simulation results in Section~\ref{simulation}.

\subsection{Specification of structural topic model hyperparameters}\label{subsec:prior}

The distribution
$p(\boldsymbol{s}^{(1:T)} \mid \boldsymbol{W}^{(1:T)}, \boldsymbol{V}^{(1:T)})$
depends on the hyperparameters $\boldsymbol{\Gamma}$, $\boldsymbol{\Lambda}$, and $\boldsymbol{\Psi}$. 
In particular, the prior distribution
$p(\boldsymbol{\theta}^{(1:T)} \mid \boldsymbol{V}^{(1:T)})$
is governed by the hyperparameter matrices $\boldsymbol{\Lambda}$ and $\boldsymbol{\Psi}$,
while the likelihood term
$p(\boldsymbol{W}^{(1:T)} \mid \boldsymbol{\theta}^{(1:T)})$
is determined by the hyperparameter matrix $\boldsymbol{\Gamma}$.
It is useful to recall that $\boldsymbol{\Lambda}$ encodes prior expectations about the contribution of covariate effects, i.e., socioeconomic indicators, in explaining the occupational group structure.
Conversely, the columns $\boldsymbol{\gamma}_k$ of $\boldsymbol{\Gamma}$ define the vocabulary \emph{rules} that characterize each topic, governing the probability of word usage across topics of interest.
In view of this, when an external corpus related to the occupational domain is available, $\boldsymbol{\Gamma}$ can be set by maximizing the likelihood of that corpus, thereby encoding prior vocabulary structure before analyzing the survey texts.
Similarly, if auxiliary data on socioeconomic indicators in occupations exist, they may be used to specify the values of $\boldsymbol{\Lambda}$, reflecting prior knowledge about occupational groupings.

In practice, when a sufficiently large period of survey responses is available, following \citet{roberts2016}, these hyperparameters can be directly estimated from the data jointly with the logistic-normal approximations of
$p(\boldsymbol{\theta}^{(1:T)} \mid \boldsymbol{W}^{(1:T)}, \boldsymbol{V}^{(1:T)})$.
Specifically, we use the textual data $\boldsymbol{W}$ and socioeconomic indicators $\boldsymbol{V}$ observed over an initial training period $1{:}t^\star$, for a certain $t^\star$ sufficiently large, and apply the non-conjugate variational Expectation–Maximization algorithm of \citet{roberts2016}, obtaining regularized point estimates $\hat{\boldsymbol{\Gamma}}_{t^\star}$, $\hat{\boldsymbol{\Lambda}}_{t^\star}$, and $\hat{\boldsymbol{\Psi}}_{t^\star}$ for the hyperparameters.
We keep the hyperparameters fixed
for all subsequent analyses, ensuring consistent topic interpretation over time.
If the analysis horizon $T$ extends far beyond $t^\star$, such that occupational terminology, the relationships between jobs and socioeconomic, or educational factors may have evolved, it is advisable to re-estimate
$\hat{\boldsymbol{\Gamma}}_{T}$, $\hat{\boldsymbol{\Lambda}}_{T}$, and $\hat{\boldsymbol{\Psi}}_{T}$ using the most recent corpus.
This ensures alignment of our prior knowledge with contemporary language use and the evolving socioeconomic structure of occupations.

\subsection{Online learning between months}
\label{sec:algo3.2}
We now formulate the two-stage approach, previously introduced, in a sequential manner to update the estimates for each new time point.
Considering the Markovian dynamics of the regression coefficients, as specified in Subsection \ref{subsec:beta-prior}, and the conditional independence of subjects within the latent Gaussian vectors $\boldsymbol{z}_{n}$ and $\boldsymbol{\theta}_{n}$ in Equations \eqref{eq:model}--\eqref{eq:tm1}, we can decompose the terms in the numerator of Equation \eqref{eq:post-dec} across time.
Specifically, given fixed hyperparameters, we have
\begin{align*}
    p(\boldsymbol{s}^{(1:T)}, \boldsymbol{B}^{(1:T)} \vert \mathcal{D}^{(1:T)}) &\propto
    \prod_{t=1}^{T} \big\{p(\boldsymbol{Y}^{(t)} \vert \boldsymbol{B}^{(t)}, \boldsymbol{s}^{(t)}, \boldsymbol{X}^{(t)}) \,p(\boldsymbol{B}^{(t)} \vert \boldsymbol{B}^{(t-1)}) \, \mathcal{L}(\boldsymbol{W}^{(t)},\boldsymbol{V}^{(t)},\boldsymbol{s}^{(t)}),
\end{align*}
 with $\mathcal{L}(\boldsymbol{W}^{(t)},\boldsymbol{V}^{(t)},\boldsymbol{s}^{(t)}) = \int_{\theta} p( \boldsymbol{W}^{(t)} \vert \boldsymbol{\theta}^{(t)}) p(\boldsymbol{s}^{(t)} \vert \boldsymbol{\theta}^{(t)}) p(\boldsymbol{\theta}^{(t)} \vert \boldsymbol{V}^{(t)}) \, \textnormal{d} \theta$ and $p(\boldsymbol{B}^{(1)} \vert \boldsymbol{B}^{(0)}) = p(\boldsymbol{B}^{(1)})$.
We can obtain 
$p(\boldsymbol{s}^{(T)}, \boldsymbol{B}^{(T)} \vert \mathcal{D}^{(1:T)})$ by simply noting
that
\begin{align*}
p(\boldsymbol{s}^{(T)}, \boldsymbol{B}^{(T)}\vert \mathcal{D}^{(1:T)})  & \propto   p(\boldsymbol{Y}^{(T)} \vert \boldsymbol{B}^{(T)}, \boldsymbol{s}^{(T)}, \boldsymbol{X}^{(T)})\, p(\boldsymbol{B}^{(T)} \vert \boldsymbol{B}^{(T-1)})   \\ 
& \times 
p(\boldsymbol{s}^{(T-1)}, \boldsymbol{B}^{(T-1)} \vert \mathcal{D}^{(1:T-1)}) \mathcal{L}(\boldsymbol{W}^{(T)},\boldsymbol{V}^{(T)},\boldsymbol{s}^{(T)}).
\end{align*}
If a sample from  
$p(\boldsymbol{B}^{(T)} \vert \boldsymbol{B}^{(T-1)}) p(\boldsymbol{s}^{(T-1)}, \boldsymbol{B}^{(T-1)} \vert \mathcal{D}^{(1:T-1)})$ is available in the form of particles; it suffices to sample from  
$\mathcal{L}(\boldsymbol{W}^{(T)},\boldsymbol{V}^{(T)},\boldsymbol{s}^{(T)}) $ 
and re-sample the pair $(\boldsymbol{B}^{(T)}, \boldsymbol{s}^{(T)})$ with weight 
$\omega^{(t)}= p(\boldsymbol{Y}^{(T)} \vert \boldsymbol{B}^{(T)}, \boldsymbol{s}^{(T)}, \boldsymbol{X}^{(T)})$ 
to obtain a Monte Carlo approximation of the \emph{filtering distribution}  
$p(\boldsymbol{s}^{(T)}, \boldsymbol{B}^{(T)} \vert \mathcal{D}^{(1:T)})$,  
which can then be used for the next time period.
The procedure is summarized in Algorithm \ref{alg:particle_filter}, which takes as input $\hat{\boldsymbol{\Gamma}}_{t^\star}$, $\hat{\boldsymbol{\Lambda}}_{t^\star}$, and $\hat{\boldsymbol{\Psi}}_{t^\star}$, the number of time periods $T$, and the number of particles $J$.

\begin{algorithm}[t]
\caption{Between Months Filtering Algorithm}
\label{alg:particle_filter}
\begin{algorithmic}
\Require Fixed $\hat{\boldsymbol{\Gamma}}_{t^\star}$, $\hat{\boldsymbol{\Lambda}}_{t^\star}$, $\hat{\boldsymbol{\Psi}}_{t^\star}$; number of time period iterations $T$; number of particles $J$; data $\mathcal{D}^{(1:T)}$.
\For{$t=1$ to $T$}
    \begin{enumerate}
        \item Draw $J$ particles from the multivariate Logistic-Normal variational distribution $p(\boldsymbol{\theta}^{(t)} \vert \boldsymbol{W}^{(t)}; \boldsymbol{V}^{(t)} )$. Denote with $\boldsymbol{\theta}^{(t)}_j$ the $j$--th particle;
        \item For each $j=1,\ldots, J$, sample $\boldsymbol{s}^{(t)}_j$ from $p(\boldsymbol{s}^{(t)} \vert \boldsymbol{\theta}_j^{(t)})$.\\
        \begin{enumerate}
        \item[\textbf{If}]
       $t=1$:       Sample $J$ particles $\boldsymbol{B}^{(1)}_j$ from $p(\boldsymbol{B}^{(1)})$.
        \item[\textbf{Else:}] \textbf{If} $t>1$:
             Sample $J$ particles $\boldsymbol{B}^{(t)}_j$ from $p(\boldsymbol{B}^{(t)} \vert \boldsymbol{B}^{(t-1)}_j)$ using previous time period particles $\boldsymbol{B}^{(t-1)}_j$.
                \end{enumerate}

        \item Compute weights: 
        $\omega^{(t)}_j =  p(\boldsymbol{Y}^{(t)} \vert \boldsymbol{B}^{(t)}_j, \boldsymbol{s}^{(t)}_j)$, 
        for each particle $(\boldsymbol{B}^{(t)}, \boldsymbol{s}^{(t)})_j$ and
         normalize them $\tilde{\omega}_j^{(t)}={\omega}_j^{(t)}/\sum_{m=1}^{J}{\omega}_m^{(t)}$; \label{line:4}
        \item Resample the $J$ particles $(\boldsymbol{s}^{(t)}, \boldsymbol{B}^{(t)})_j$ according to the corresponding weights $\tilde{\omega}_j^{(t)}$. \label{line:5}
        \item Relabel the group indices in $(\boldsymbol{s}^{(t)}_j, \boldsymbol{B}^{(t)}_j)$. \label{line:6}
    \end{enumerate}
\EndFor
\end{algorithmic}
\end{algorithm}

The Steps \ref{line:4} and \ref{line:5} in Algorithm \ref{alg:particle_filter} require updating a multidimensional parameter that also depends on the number of units observed in each time period.  
As widely recognized in the literature \citep{kantas2015particle, doucet2000sequential}, updating static multidimensional parameters often leads to particle degeneracy, resulting in poor samples characterized by a limited number of effective particles.  
To mitigate this issue, we introduce a set of adjustments, which are detailed in Subsection \ref{subsec:seq-mc}.

To maintain consistency in group membership labels across particles, in Step \ref{line:6} of Algorithm \ref{alg:particle_filter} we apply the relabeling algorithm of \citet{stephens2000} on $\boldsymbol{s}^{(t)}$, as implemented in the {\sf R} package \texttt{mcclust} \citep{fritsch2022}. Based on the resulting relabeling, we consistently align the group indices $k = 1, \ldots, K$ in each particle of $\boldsymbol{B}^{(t)}$.

\subsection{Within-month update and parallel computing}
\label{subsec:seq-mc}

Let us consider a specific iteration of Algorithm~\ref{alg:particle_filter} corresponding to time period $t$, focusing particularly on Steps \ref{line:4} and \ref{line:5}.  
Since  
$\boldsymbol{s}^{(t)}$ is an $N_t$-dimensional vector and $\boldsymbol{B}^{(t)}$ is a $K \times P \times Q$-dimensional array,  
the parameter space becomes high-dimensional when $K$, $Q$, or even just the number of observations $N_t$ within the period is large.  
As is well known in the literature, the \emph{curse of dimensionality} causes particles to be widely dispersed,  
leading to only a handful of normalized weights $\tilde{\omega}_j^{(t)}$ with significantly higher values than the others.  
This results in the resampling of a small number of particles in Step \ref{line:5} of Algorithm~\ref{alg:particle_filter},  
making inference on $(\boldsymbol{s}^{(t)}, \boldsymbol{B}^{(t)})$ infeasible.  
To address this issue, we replace Steps \ref{line:4} and \ref{line:5} with a filtering procedure that sequentially updates the parameter $(\boldsymbol{s}^{(t)}, \boldsymbol{B}^{(t)})$  
by incorporating one observation at a time within the month.  
To mitigate particle degeneracy in the marginal posterior of $\boldsymbol{B}^{(t)}$,  
we use a \emph{rejuvenation step} in the algorithm,  
as is typically done in SMC$^{2}$ algorithms \citep{chopin2013smc2}.  
The full procedure is detailed in Algorithm~\ref{alg:particle_filter2}; here we focus on the main ideas.

\begin{algorithm}[t]
\caption{Within-Month Particle Filtering Algorithm with Rejuvenation}
\label{alg:particle_filter2}

\begin{algorithmic}

\Require $J$-size sample of particles $(\boldsymbol{s}, \boldsymbol{B})_j$, with $j=1,\ldots,J$.

\For{$n=1$ to $N_t$}
    \begin{enumerate}[I.]
        \item Compute weights: $\omega_{nj} =  p(\boldsymbol{y}_n \vert \boldsymbol{B}^{(t)}_j, s^{(t)}_{nj})$, 
        for each particle $(\boldsymbol{B}^{(t)}_j, s^{(t)}_{nj})$ and
         normalize them $\tilde{\omega}_{nj} =\omega_{nj} /\sum_{m=1}^{J}{\omega}_{nm} $; \label{line:12}
        \item Resample the $J$ particles $\boldsymbol{B}^{(t)}_j$ according to the corresponding weights $\tilde{\omega}_j^{(t)}$, for $j=1,\ldots,J$. \label{line:22}
        \item Rejuvenate the particle sample $\boldsymbol{B}^{(t)}_j$ ($j=1,\ldots,J)$, following the steps below. \label{line:32} \newline        
        \textbf{for} $j=1$ to $J$ \textbf{do}
            \begin{enumerate}[i.]
                \item Define the proposal coefficients $\tilde{\boldsymbol{B}}^{(t)}_j$: \newline
                \textbf{for} $k \neq s_{nj}$ \textbf{do} $\tilde{\boldsymbol{B}}^{(t)}_{kj} = \boldsymbol{B}^{(t)}_{kj}$ \textbf{end for}; \newline
                \textbf{for} $k = s_{nj}$ \textbf{do}
                  sample $\tilde{\boldsymbol{B}}^{(t)}_{kj}$ from the prior $p(\boldsymbol{B}_{k}^{(1)})$ or jitter $\boldsymbol{B}^{(t)}_{kj}$ \textbf{end for}.
                \item Compute the acceptance ratio \label{line:322}
                \begin{equation*}
                    \alpha = \frac{ p(\tilde{\boldsymbol{B}}^{(t)}_{j} \vert \boldsymbol{Y}^{(1:t-1)}, \boldsymbol{y}_{1},\ldots,\boldsymbol{y}_{n}, \boldsymbol{X}^{(1:t)}, \boldsymbol{W}^{(1:t)}, \boldsymbol{V}^{(1:t)})
                    }{
                     p(\boldsymbol{B}^{(t)}_{j} \vert \boldsymbol{Y}^{(1:t-1)}, \boldsymbol{y}_{1},\ldots,\boldsymbol{y}_{n},\boldsymbol{X}^{(1:t)}, \boldsymbol{W}^{(1:t)}, \boldsymbol{V}^{(1:t)})
                    }
                    \frac{p(\boldsymbol{B}^{(t)}_{j} \vert \tilde{\boldsymbol{B}}^{(t)}_{j})}{
                    p(\tilde{\boldsymbol{B}}^{(t)}_{j} \vert \boldsymbol{B}^{(t)}_{j})
                    }
                \end{equation*}
                \item Sample $u$ from uniform distribution $\mathcal{U}(0,1)$. \newline
                \textbf{If} $u\leq \alpha$: set $\boldsymbol{B}^{(t)}_{kj}=\tilde{\boldsymbol{B}}^{(t)}_{kj}$
            \end{enumerate}
        \textbf{end for}
    \end{enumerate}
\EndFor

\noindent Update the particle sample $\boldsymbol{s}^{(t)}_j$ ($j=1,\ldots,J$) by sampling $\boldsymbol{\theta}^{(t)}_j$ from
    $p(\boldsymbol{\theta}^{(t)}_j \vert \boldsymbol{B}^{(t)}_{kj}, Y^{(t)}, X^{(t)})$ and then $\boldsymbol{s}^{(t)}_j$ from $p(\boldsymbol{s}^{(t)}_j \vert \boldsymbol{\theta}^{(t)})$.
\end{algorithmic}
\end{algorithm}

Assume that a sample of size $J$ of $(\boldsymbol{s}^{(t)}, \boldsymbol{B}^{(t)})$ is available from the prior distribution  
$p(\boldsymbol{s}^{(t)} \vert \boldsymbol{W}^{(t)}, \boldsymbol{V}^{(t)}) 
p(\boldsymbol{B}^{(t)} \vert \boldsymbol{B}^{(t-1)})$. For $n = 1, \dots, N_t$, observations $(\boldsymbol{y}_n^{(t)}, \boldsymbol{x}_n^{(t)})$ are processed sequentially to update posterior beliefs 
on group memberships $\boldsymbol{s}^{(t)}$ and regression coefficients $\boldsymbol{B}^{(t)}$.

In Steps~\ref{line:12}--\ref{line:22} of Algorithm~\ref{alg:particle_filter2}, particle weights are computed from the likelihood contribution of the current observation and used to resample the particle set.
While this procedure yields a valid filtering distribution, the finite particle sample can lead to a rapid depletion of diversity in $\boldsymbol{B}^{(t)}$ as new observations are incorporated, particularly when the number of groups $K$ is large.
This phenomenon is known and commonly discussed in SMC literature \citep[see, for example,][]{chopin2020introduction, kantas2015particle}.

We address particle depletion by incorporating a Metropolis-Hastings rejuvenation step \citep{chopin2013smc2} after each filtered observation. 
This step allows particles to explore the local posterior distribution of $\boldsymbol{B}^{(t)}$ and helps restore diversity in the particle population.
The acceptance ratio $\alpha$ used in Step~\ref{line:322} of Algorithm~\ref{alg:particle_filter2} is described in Appendix~C of the SM.

New proposal values $\tilde{\boldsymbol{B}}^{(t)}_{j}$ are generated either from the prior distribution or by jittering the current particle values.  
The prior density $p(\boldsymbol{B}^{(t)}_{j} \vert \mathcal{D}^{(1:t-1)})$ is evaluated at the current state $\boldsymbol{B}^{(t)}_{j}$ using results from the previous iteration $n-1$, whereas its value at the proposal $\tilde{\boldsymbol{B}}^{(t)}_{j}$ is computed via numerical integration. 
Computational details on the evaluation of the prior density and on the acceptance ratio used in Step~\ref{line:322} of Algorithm~\ref{alg:particle_filter2} are provided in Appendix~C of the SM.

The algorithm admits partial parallelization.  
One strategy may be to execute Step~\ref{line:32} in parallel for $j = 1, \ldots, J$.  
However, this strategy requires calculating Steps~\ref{line:12}-\ref{line:22} for all particles, which should number in the thousands to ensure adequate approximation performance.  
To reduce computational demand while enhancing parameter space exploration, we propose an alternative strategy in which we run multiple independent parallel instances of Algorithm~\ref{alg:particle_filter2}, each using a small to moderate number of particles $J$.  
Within each instance, the order of observations in $\mathcal{D}^{(t)}$ is randomly permuted, leveraging the order-invariant model specification within each time period $t$.  This strategy improves mixing and reduces sensitivity to local modes while maintaining computational efficiency.
To promote effective exploration of the parameter space at the first time period, we approximate the filtering distribution
$p(\boldsymbol{s}^{(1)}, \boldsymbol{B}^{(1)} \mid \mathcal{D}^{(1)})$
using HMC implemented in {\sf Stan}.

At each time period $t$, one could instead rely on MCMC-based posterior approximation schemes, such as those proposed in \cite{goudie2019joining}.
While these methods can improve exploration of the parameter space, they typically require running MCMC until convergence at each time period, which substantially increases computational cost.
Our proposal can therefore be viewed as a more computationally efficient alternative, as demonstrated by the simulation studies in Section~\ref{simulation}.


\section{Algorithm assessment via simulation experiments}
\label{simulation}

We evaluate the performance of the proposed \textsc{SMC} algorithm by comparing it with several alternatives. Specifically, we examine its ability to approximate the posterior predictive distribution
$p(\tilde{\boldsymbol{y}}_n \mid {\boldsymbol{x}}_n, \mathcal{D}^{(1:t)})$
for a new subject, given the availability of a sample from the Logistic-Normal distribution
$p(\boldsymbol{\theta}^{(1:t)} \mid \boldsymbol{W}^{(1:t)}, \boldsymbol{V}^{(1:t)})$. 

We mimic the data-generating process of the motivating application described earlier. 
For two consecutive periods ($t \in \{1,2\}$), we consider a $P = 4$-dimensional ordinal response variable, similar in structure to the four SNAP variables of the application.
We design eight scenarios by varying $(N_1, N_2) \in \{(200, 250), (1000, 1200)\}$, $(Q, K) \in \{(3, 5),(5, 10)\}$, and a sparsity index $\varsigma \in \{0.2, 0.4\}$ used to control the amount of regression coefficients set to zero.
In each scenario, we first randomly set the binary covariate matrix $\boldsymbol{X}$, the regression coefficients $\bar{\boldsymbol{B}}^{(1)}$ and $\bar{\boldsymbol{B}}^{(2)}$, and the true clustering belonging $\bar{\boldsymbol{s}}$.
Then, we replicate for $R = 20$ times the generation of response variables $\boldsymbol{Y}^{(1)}$ and $\boldsymbol{Y}^{(2)}$ according to the ordinal multivariate mixture model in Equation~\eqref{eq:model}. All simulations were run in parallel on a Windows machine equipped with an AMD EPYC~7413~40-Core Processor (2.65~GHz), using 30~cores.
Full details of simulation settings are provided in Appendix~D of the SM.

For each replication, different estimation strategies are adopted to approximate the posterior predictive distribution $p(\boldsymbol{y}_n \mid \boldsymbol{x}_n, \mathcal{D}^{(1:2)})$, for the subjects observed in the second period, i.e., $n = N_1,\ldots, N$, with $N=N_1+N_2$, conditional on the full observed dataset $(\boldsymbol{X}^{(1:2)}, \boldsymbol{Y}^{(1:2)})$ and given
the availability of a sample from a Logistic-Normal distribution $p(\boldsymbol{\theta}^{(1:2)}\vert \boldsymbol{W}^{(1:2)}, \boldsymbol{V}^{(1:2)}).$
For all competing algorithms, the Logistic-Normal distribution hyperparameters used to sample the \textit{true} $\boldsymbol{\theta}_n$, for $n=1,\ldots, N$ are available. 
The topic model results are provided as a prior distribution for clustering, i.e., the comparisons refer to the ability to correctly update the prior belief on $\boldsymbol{B}^{(2)}$, and $\boldsymbol{s}$ after observing $\boldsymbol{X}^{(1:2)}$ and $\boldsymbol{Y}^{(1:2)}$.

As a gold-standard benchmark, we consider the predictive distribution sample obtained from 
HMC run in {\sf Stan} \citep{carpenter2017stan}. 
We denote this benchmark as $p_{\textsc{HMCFull}}(\boldsymbol{y}_n \mid \boldsymbol{x}_n, \mathcal{D}^{(1:2)})$.  
This method does not allow for online updates and requires the full dataset from the entire period observed to estimate jointly $\boldsymbol{B}^{(1)}$, $\boldsymbol{B}^{(2)}$, and $\boldsymbol{\theta}_n$ for all $n = 1,\ldots, N$. 
This strategy offers good sampling qualities, but it could be time and memory-consuming, especially in higher dimensions (may be them $Q$, $T$, $N_t$, or $K$). 
In the simulations, the \textsc{HMCFull} is run using $30$ parallel chains with $2000$ iterations each, discarding the first $1850$ iterations as warm-up, leading to a sample with $4500$. 

We use $p_{\textsc{HMCFull}}(\boldsymbol{y}_n   \vert x_n, \mathcal{D}^{(1:2)})$ to assess the quality of approximation of alternative algorithms.
As evaluation metric, we consider the empirical Kullback–Leibler divergence between $p_{\textsc{HMCFull}}(\boldsymbol{y}_n  \vert x_n, \mathcal{D}^{(1:2)})$ and the predictive distributions of alternative algorithms, averaged across all subjects in period $t=2$. 
We rely on the implementation available in the {\sf R} package \texttt{kldest} \citep{hartung2025}, with divergence close to zero indicating that the method closely approximates the full HMC predictive distribution. 

As a first approach, we consider the predictive distribution approximated by our proposed method, described in Algorithms~\ref{alg:particle_filter}–\ref{alg:particle_filter2}, which generates $J = 4500$ particles to approximate the posterior predictive distribution at $t = 2$.
Following the recommendations in Subsection~\ref{subsec:seq-mc}, we employ parallel computing by running 30 parallel instances with 150 particles each.
The resulting distribution is denoted as $p_{\textsc{SMC}}(\boldsymbol{y}_n \mid \boldsymbol{x}_n, \mathcal{D}^{(1:2)})$.
We compare our results against two types of competitors: (i) Laplace approximation methods using the full sample, which are known for their computational efficiency, and (ii) sequential sampling-based methods, which typically offer higher approximation accuracy.
The Laplace approximation based on the full periods is denoted by $p_{\textsc{LFull}}(\boldsymbol{y}_n \mid \boldsymbol{x}n, \mathcal{D}^{(1:2)})$.
We obtain this distribution using the {\sf cmdstanr} implementation of the Laplace approximation algorithm, applied to the same model and data used for $p_{\textsc{HMCFull}}(\boldsymbol{y}_n \mid \boldsymbol{x}_n, \mathcal{D}^{(1:2)})$.
This implementation jointly estimates the parameters $\boldsymbol{B}^{(1)}$, $\boldsymbol{B}^{(2)}$, and $\boldsymbol{\theta}_n$ for all $n = N_1,\ldots, N$, used to get $4{,}500$ draws from the respective posterior predictive distribution.
The remaining two competitors enable sequential inference through a two-step procedure.
In the first step, we fit a flexible mixture model to the posterior samples of $\boldsymbol{B}^{(1)}$, obtaining a closed-form approximation of $p(\boldsymbol{B}^{(1)} \mid \mathcal{D}^{(1)})$.
This approximation, combined with the dynamic specification in Equation~\eqref{eq:beta-prior}, provides an approximate prior for $\boldsymbol{B}^{(2)}$ conditional on $\boldsymbol{B}^{(2)}$ conditional on first period data $ \mathcal{D}^{(1)}$.
In the second step, we estimate the predictive distribution at $t = 2$ using either a {\sf Stan} implementation of HMC or the Laplace approximation algorithm, obtaining $4{,}500$ draws from $p_{\textsc{HMC2Step}}(\boldsymbol{y}_n \mid \boldsymbol{x}n, \mathcal{D}^{(1:2)})$ and $p{\textsc{L2Step}}(\boldsymbol{y}_n \mid \boldsymbol{x}_n, \mathcal{D}^{(1:2)})$, respectively, for $n = N_1,\ldots, N$.
These two approaches represent computationally efficient alternatives to the full-sample algorithms, as they require an approximation of $p(\boldsymbol{B}^{(t)} \mid \mathcal{D}^{(1:(t-1))})$ as input, while updating only the parameters at time $t$ (in this case, $t = 2$) like our proposed method.

\begin{figure}[ht]
    \centering
    \begin{minipage}{0.49\linewidth}
        \centering
        \includegraphics[width=\linewidth]{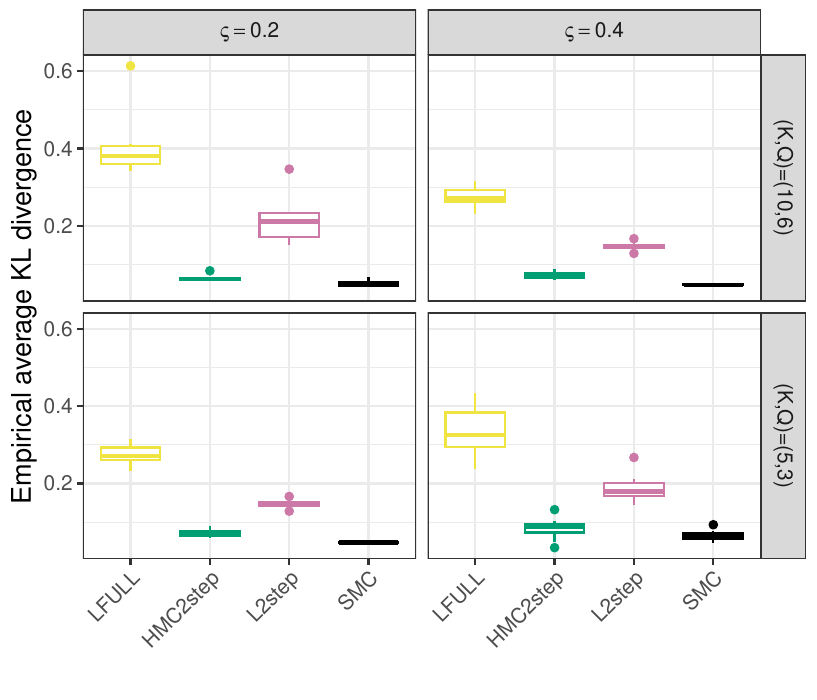}
    \end{minipage}
    \hfill
    \begin{minipage}{0.49\linewidth}
        \centering
        \includegraphics[width=\linewidth]{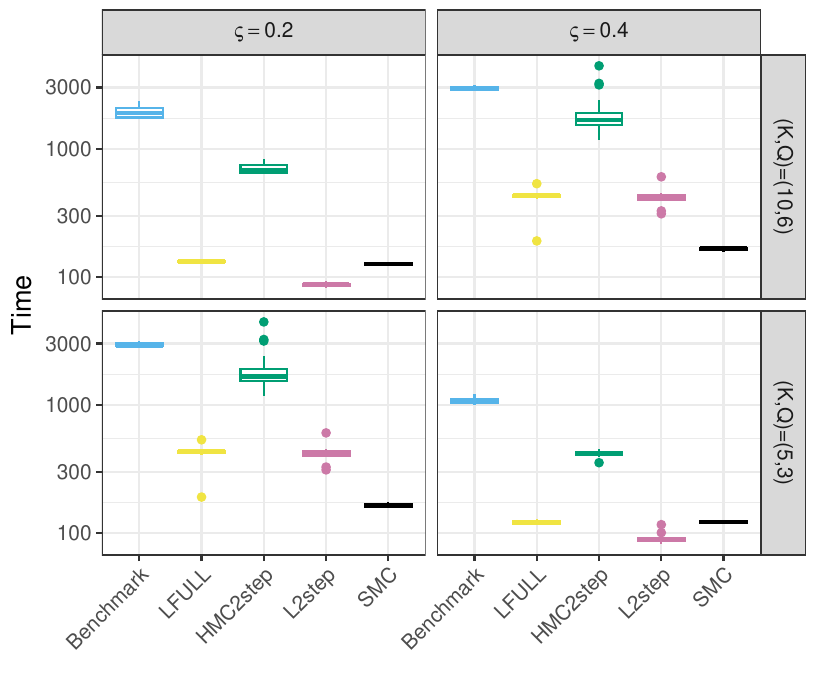}
    \end{minipage}

    \caption{
    Boxplots of the distribution over $R = 20$ simulation replications of the average KL divergence between the benchmark predictive posterior samples provided by \textsc{HMCFull} and those generated by the competing models (left), and of the time (in seconds with axis in base-10 logarithmic scale) required to generate predictive posterior samples for subjects at $t = 2$ for each competing model (right), evaluated across varying $\varsigma$, $K$, and $Q$.}
    \label{fig:sim}
\end{figure}

Simulation results assessing the quality of the predictive approximations are reported in the left panel of Figure~\ref{fig:sim}.
As expected, Laplace-based methods exhibit the largest discrepancies, likely due to their limited ability to capture mixture structures. In contrast, the SMC and HMC two-step methods provide comparable approximation quality, with our proposed approach showing slightly better performance.
This advantage appears more systematic as the number of nonzero parameters increases, that is, in scenarios with larger $(K, Q)$ values or smaller $\varsigma$.

Other than providing better quality in approximation, the proposed method also offers a significant advantage in terms of computational efficiency. 
The right panel of Figure~\ref{fig:sim} shows the time (in seconds) required by each algorithm to generate predictive samples for $n = N_1, \ldots, N$.
The time required to estimate \(\boldsymbol{B}^{(1)}\) is omitted whenever possible; however, for the non-sequential algorithms (HMCFULL and LFULL) this step cannot be isolated, so the reported times include the joint estimation of \(\boldsymbol{B}^{(1)}\) and \(\boldsymbol{B}^{(2)}\). This does not significantly disadvantage the full-sample methods, given the limited number of periods (\(T = 2\)).  
In fact, the advantage of sequential approaches emerges as \(T\) increases: they update previously estimated quantities using new data, leading to a computational cost that grows roughly linearly in \(T\), whereas full-sample methods must reprocess all past data, causing their burden to escalate quickly.  
Our SMC approach demonstrates substantial computational savings relative to Hamiltonian Monte Carlo methods, primarily because it avoids the costly warm-up phases.
Furthermore, its fully parallelizable implementation enables performance comparable to faster, though less accurate, alternatives such as the Laplace approximations.
Notably, this remains true even in highly parameterized settings, where the SMC algorithm continues to perform competitively in both speed and accuracy.

Additional simulation scenarios, yielding consistent conclusions across settings with smaller $N$, are reported in Figures~S3--S4 in Appendix D of the SM.

\section{Behavioral risk factors in Italy}
\label{application}

\subsection{Data}

We analyze data from the Italian health surveillance system PASSI. The population of reference is Italian adults ages 18 to 69. For additional information, please refer to \citet{baldissera2011peer} and the \cite{epicentro_passi} website. We analyze data from 2014 to 2023, as job-related information has been available in the survey since 2014.
Approximately $35\%$ of the observations correspond to unemployed respondents, who are excluded from the analysis, yielding a final sample of 133{,}485 observations.
The variables used in the model of Figure~\ref{fig:model-dag}, i.e., the demographic covariates $\boldsymbol{x}_n$, socioeconomic characteristics $\boldsymbol{v}_n$, and behavioral outcomes $\boldsymbol{y}_n$, are defined in Table~\ref{tab:data}.
Thresholds $\tau_{pc}$ were set according to the number of categories of each ordinal outcome, so that the corresponding cumulative probabilities of a standard normal distribution are approximately balanced between adjacent thresholds. Specifically, we used
$(\tau_{p1},\tau_{p2},\tau_{p3},\tau_{p4},\tau_{p5})= (-1, -0.5, 0, 0.5,1)$ for smoking, 
$(\tau_{p1},\tau_{p2},\tau_{p3})= (-0.75, 0, 0.75)$ for nutrition and alcohol consumption, and $(\tau_{p1},\tau_{p2})= (-0.5,  0.5)$ for physical inactivity. The classification of smoking and alcohol consumption followed the guidelines of \cite{epicentro_passi}. 
In contrast, the indicators for nutrition and physical activity were retained as originally recorded in the questionnaire.

\begin{table}[]
    \centering
\begin{tabular}{ll|p{0.73\linewidth}}
    \toprule
    &\textbf{Variable}     &  
    \textbf{Description}\\
    \midrule
     $\boldsymbol{x}_n$&Sex   &  Male; Female\\
     &Age & [18,30); [30, 50); [50, 69] \\
     &Macro Area & Center; Islands; North-East; North-West; South \\
     \midrule
     $\boldsymbol{v}_n$&Educational level & Low: if below high school; high: otherwise\\
     &Economic problem & 0: if the respondent easily meets financial needs; 1: otherwise\\
          \midrule
     $\boldsymbol{y}_n$& Smoke & 0: never smoked, 1: currently smokes less than a cigarette, 2: currently smokes 1-4 cigarettes, 3: currently smokes 5-9 cigarettes, 4: currently smokes 10-19 cigarettes, 5: currently smokes at least 20 cigarettes (per day)\\
     &Nutrition & 3: No fruit; 2: 1-2 fruit portions; 1: 3-4 fruit portions; 0: 5+ fruit portions (per day)\\
     &Alcohol & Sum of three binary indicators: high alcohol consumption (more than two drinks per day for men and more than one drink per day for women), drinking outside meals, and at least one episode of binge drinking, where a value of 1 indicates the presence of risky behavior. The variable takes values in the set $\{0,1,2,3\}$. \\
     &Physical activity & 0: Intense; 1: moderate; 2: no physical activity (in the last 30 days).\\
     \bottomrule
    \end{tabular}
    \caption{Description of the variables included in the analysis.} 
    \label{tab:data}
\end{table}

For the textual data about occupation, we examine three variables related to the respondents' employment. The first stems from the query: ``Can you tell me what you do for a living?'', i.e., an open-ended question. The second involves the classification of the declared job according to ISTAT coding \citep{istat_cp2011}, while the third one delineates nineteen occupational sectors defined both in Tables~S1-S2 of Appendix A in the SM.
While the latter two provide structured categorical information, they may fail to capture the rich contextual and semantic nuances contained in the open-ended responses.
For each respondent $n$, we therefore integrate all three occupational sources into a unified textual description of the job. After removing stopwords, we construct a document–term matrix containing all unigrams appearing in these fields. 
Because many occupations are repeated, we retain only the unique combinations of texts and $\boldsymbol{v}_n$ for the estimation of the STM, ensuring that two respondents with identical occupational descriptions $\boldsymbol{w}_n$ and socioeconomic characteristics $\boldsymbol{v}_n$ share the same distribution $\Pr(\boldsymbol{\theta_n} \vert \boldsymbol{W}, \boldsymbol{V})$. This does not preclude assigning different group proportions to these respondents after observing the behavioral data $\boldsymbol{Y}$.

Following \citet{roberts2016}, we selected the STM topic number by fitting models with $K=15,\ldots,35$ and comparing their held-out likelihood and semantic coherence. 
The model with $K=30$ topics was chosen for its strong interpretability-performance trade-off and as part of a deliberate overparameterization strategy. 
Using a relatively large number of topics helps capture fine-grained occupational and socioeconomic nuances, yielding a flexible representation of occupational heterogeneity.
At the same time, the shrinking non-local prior regularizes the mixture model by automatically down-weighting redundant or weakly identified topics. 
This allows the model to capture genuine variation while merging topics with similar behavioral patterns, producing a parsimonious yet nuanced view of occupational groups.
We train the STM using data collected up to December~2019, corresponding to  $t^{\star} = 72$,  keeping fixed $\hat{\boldsymbol{\Gamma}}_{t^\star}$, $\hat{\boldsymbol{\Lambda}}_{t^\star}$, and $\hat{\boldsymbol{\Psi}}_{t^\star}$ for all $t=1,\ldots,120$.
Further details on prior specification and computational aspects are provided in Appendix~E of the SM.

\subsection{Results}

The algorithms presented in Section~\ref{sec:algo} allow us to investigate the filtering distribution $p(\boldsymbol{s}^{(t)}, \boldsymbol{B}^{(t)} \mid \mathcal{D}^{(1:t)})$, for any $t= 1,\ldots, T$ of interest. 
This joint distribution captures how respondents are grouped, through the posterior of $\boldsymbol{s}^{(t)}$, based on both prior information from occupational descriptions $\boldsymbol{W}^{(t)}$ and observed behavioral profiles derived from the SNAP observations $\boldsymbol{Y}^{(t)}$.
Jointly examining $\boldsymbol{s}^{(t)}$ and $\boldsymbol{B}^{(t)}$ provides insights into how identified groups evolve and differ in their behavioral patterns.

For instance, Figure~S5 in the SM shows the posterior distributions of the coefficients $\beta^{(1)}_{1k3}$, for $k = 1, \ldots, K$, highlighting the heterogeneity across latent groups in the effect of belonging to the $(50,69]$ age class on smoking-related risky behavior.
At $t = 1$, it is already evident that being an older worker increases the probability of smoking among skilled metal-industry workers, while the opposite effect is observed for farmers and agricultural workers.
Although this representation underscores the flexibility of the mixture structure in capturing heterogeneous effects across groups, it is limited in conveying temporal evolution.
Moreover, the dimensionality of the parameters is extremely high (depending on $P$, $Q$, $K$, and $N_t$), requiring careful consideration in summary representation and visualization.

As a more interpretable alternative, we focus on the \emph{filtered predictive distribution} for a \emph{profile}, i.e., a hypothetical subject with index $\tilde{n}$, unobserved in the sample but characterized by their own job text description and sociodemographic features.
Focusing on predictive distributions of profiles rather than latent parameters offers several advantages.
First, the predictive distribution is an invariant posterior summary that inherently accounts for the relative weights of joint realizations of $(\boldsymbol{s}^{(t)}, \boldsymbol{B}^{(t)})$.
More importantly, it shifts attention to the probability scale of observed responses, allowing a direct and interpretable assessment of the probabilities of risky behaviors and their temporal evolution as new data become available.
For a profile, we define the target distribution as
\begin{align*}
p(\boldsymbol{y}_{\tilde{n}} \mid \boldsymbol{x}_{\tilde{n}}, \mathcal{D}^{(1:t)}) & =  \int p(\boldsymbol{y}_{\tilde{n}} \mid \boldsymbol{x}_{\tilde{n}}, s_{\tilde{n}}, \boldsymbol{B}^{(t)}) \, \textnormal{d} p(s_{\tilde{n}},\boldsymbol{s}^{(t)},  \boldsymbol{B}^{(t)} \mid \boldsymbol{w}_{\tilde{n}}, \boldsymbol{v}_{\tilde{n}}, \mathcal{D}^{(1:t)}) 
\end{align*}
where the hypothetical subject \(\tilde{n}\) may belong to any period of interest \(t = 1, \ldots, T\).

In this framework, any individual of interest can, in principle, be monitored. 
However, to illustrate the key features of the model, it is useful to select profiles that represent specific socioeconomic environments, i.e., choosing particular values of \(\boldsymbol{w}_{\tilde{n}}\) and \(\boldsymbol{v}_{\tilde{n}}\), that emphasize the latent relationships within the data. 
We identify such representative profiles using the \texttt{findThoughts} function from the \texttt{stm} package \citep{stm-r}, which retrieves the occupational and socioeconomic profiles most strongly associated with a given topic. 
Although the socioeconomic environment is selected from observed individuals, our analysis focuses on profiles whose demographic characteristics \(\boldsymbol{x}_{\tilde{n}}\) are varied to explore how the predictive distribution of responses changes accordingly. 
Understanding how the influence of demographic covariates varies across socioeconomic contexts is crucial for both sociological interpretation and policy design. 
Through this framework, our model provides quantitative assessments of how sociodemographic factors shape the predicted probabilities of risky behaviors, together with their associated uncertainty.

\begin{figure}[ht]
    \centering
    \includegraphics[width=.8\linewidth]{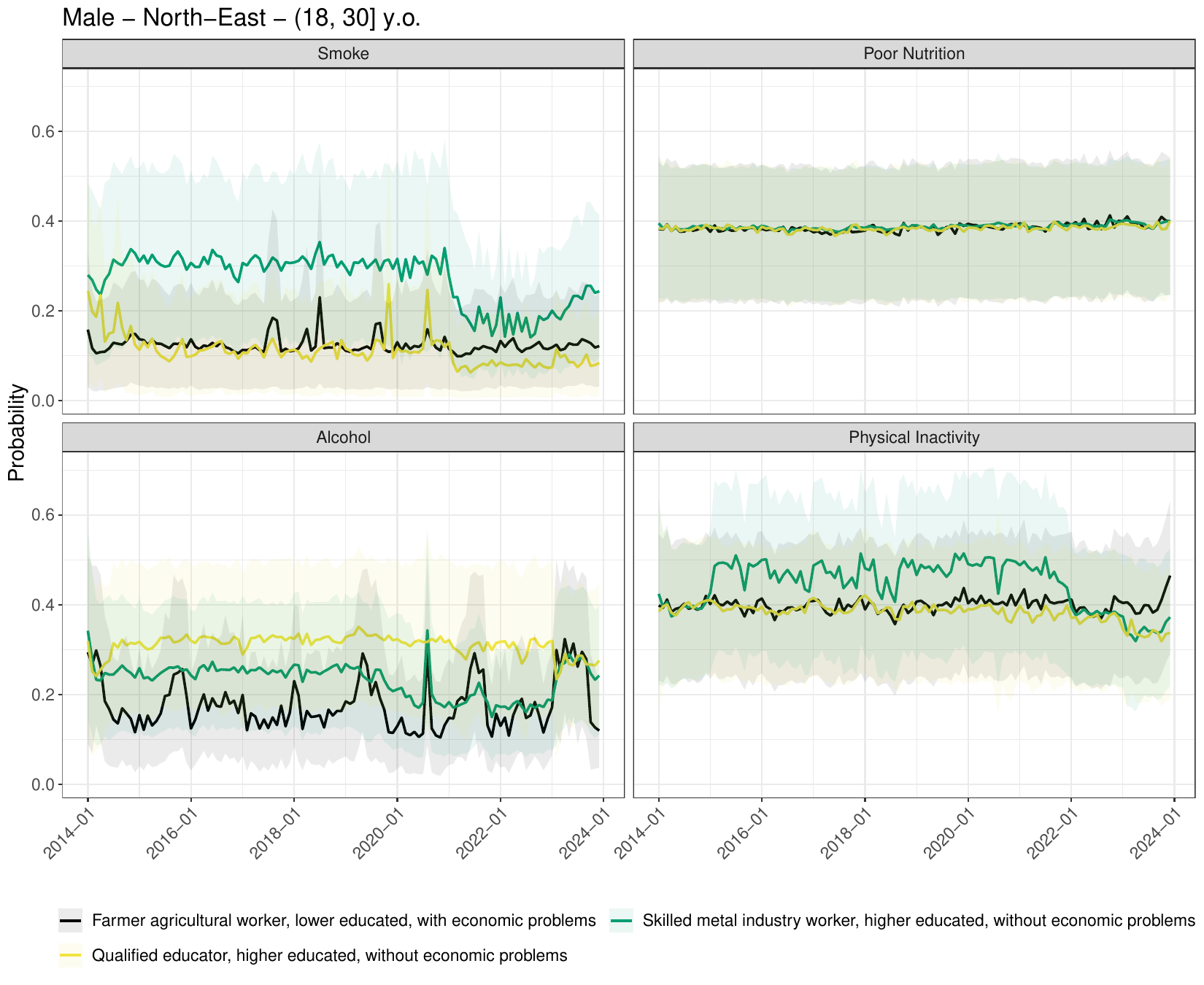} 
    \caption{Estimated risk probabilities by year for each SNAP behavior, for 18–30-year-old males from north-east Italy and three latent job profiles. The shaded areas represent the $90\%$ credible intervals.}
    \label{fig:ts}
\end{figure}

To provide an example, we present in Figure~\ref{fig:ts} the temporal dynamics of the estimated probabilities of risky behaviors for three representative profiles of young male individuals from north-east Italy, where, according to the ordinal probit specification, probabilities of risky behaviors are evaluated at the cutpoints $0$ for smoking and alcohol (i.e., starting to smoke and exhibiting at least one risky alcohol-related behavior respectively), $0.5$ for nutrition (i.e., consuming fewer than five portions of fruit per day), and $0.75$ for physical activity (i.e., being sedentary). 
We can note that before the COVID-19 pandemic, skilled metal industry workers exhibited higher smoking propensities, which later converged to the levels observed for qualified educators and unskilled agricultural workers, displaying comparable post-pandemic dynamics.
Instead, qualified educators have a higher likelihood of alcohol consumption, while skilled metal industry workers tend to be less physically active.

\begin{figure}[ht]
    \centering
    \includegraphics[width=.8\linewidth]{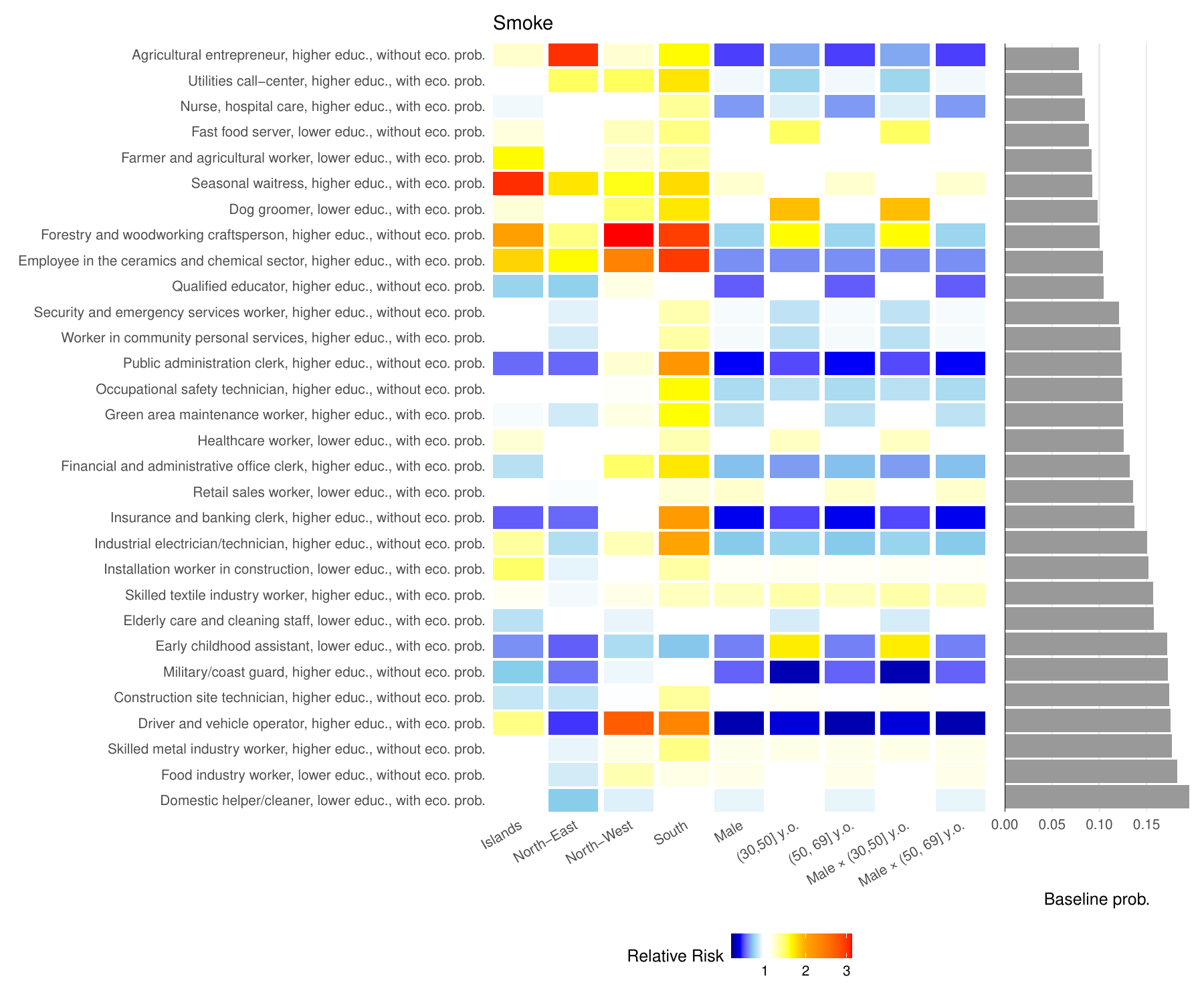}
    \caption{Smoking relative risk for each job group and socio-demographic covariates for December 2023.
    Reference individual: [18,30] female from the center macro area.}
    \label{fig:heatmap_smoke}
\end{figure}

Clearly, it is just sufficient to consider a specific $t$ to focus on a cross-sectional perspective. 
We do this by considering the relative risk, defined here as $ p_{gp}^t /  p_{\cdot p}^t$, where  $p_{\cdot p}^t$ represents the prevalence in risky behavior $p$ of a baseline group.
Such relative risks are represented in Figure~\ref{fig:heatmap_smoke}, where prevalences of the baseline categories are computed by considering the strata of $[18,30]$-year-old females living in the center macro area marginal with the same occupation characteristics. The influence of territorial context appears to differ across occupational categories, while other age and sex factors display more stable and unidirectional effects on risky behaviors, with some exceptions.
Being from the South emerges as a risk factor for smoking propensity across most hypothetical individuals, particularly among those with higher education levels.
Conversely, being male appears to be a protective factor for nearly all occupations, except for those related to industry.
These protective effects appear to be more pronounced for occupations characterized by a higher education and better economic conditions. 

These findings suggest that the relationship between socioeconomic status and SNAP behaviors is not straightforward, as not often recognized in the epidemiological literature \citep{pearce2006complexity, cao2023smoking}. 
This complexity is, however, well recognized in clinical medicine and in behavioral and medical sociology, where the interplay between individuals and their environments is understood as inherently non-trivial \citep{jama_complexity, adler1999socioeconomic, stival2025communicatingcomplexstatisticalmodels}. 
Structural conditions, micro-environmental contexts, and contextual resources can exert both protective and risk-enhancing effects, shaping health-related behaviors in ways that are far from simple. Vulnerability does not uniformly translate into unhealthy behaviors, and lifestyle factors, including occupational context, also shape such patterns \citep{BRAVEMAN2011S58}.

\begin{figure}[ht]
    \centering
    \includegraphics[width=.925\linewidth]{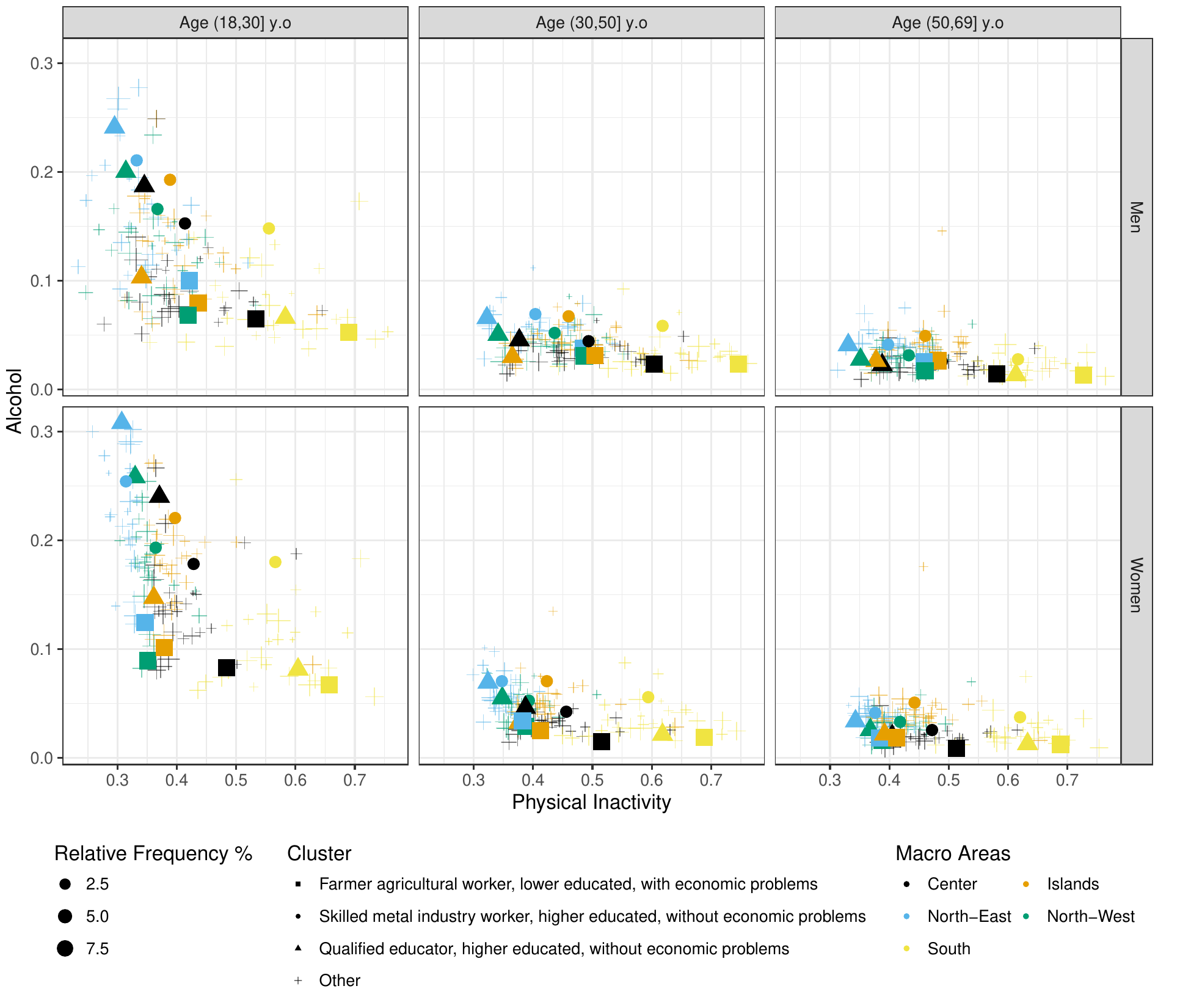}
    \caption{Estimated probabilities of engaging in risky alcohol consumption and physical inactivity, by sex, age group, macro area, and three latent job groups.}
    \label{fig:pm1}
\end{figure}

The model also permits joint analyses of behaviors.
Figure~\ref{fig:pm1} presents the posterior risk probabilities for alcohol consumption and physical inactivity across demographic covariates.
In the South (colored in yellow), physical inactivity is pronounced among unskilled agricultural workers, though it is also evident in other occupational groups.
Skilled industry workers exhibit a relatively high propensity for alcohol consumption.
Conversely, qualified educators in the North-East display higher alcohol consumption tendencies, especially among younger individuals, as also shown in Figure~S7 in the SM.
From the perspective of risky behaviors, it is clear that hazardous alcohol consumption is more pronounced among younger age groups overall.
While this can be partly explained by the different lifestyles people adopt at various stages of life, it is interesting to note that differences also emerge between sexes. 
The analyses suggest a tendency toward a higher risk of alcohol consumption among young women compared to men, a pattern particularly pronounced in the north-eastern regions of the country. In the literature, this association is well documented among adolescents \citep{hoots2023alcohol, foresta2021fumo}, and more recently it has also been observed among young women in Italy, coherently with our study \citep{messina2021knowledge}, and in the US \citep{white2020}.

These findings highlight the importance of policymakers developing targeted health promotion strategies. 
Interventions could prioritize increasing physical activity in the southern regions via investments in accessible public spaces and infrastructure, workplace wellness programs, and community-based initiatives encouraging adult participation in sports or active transport. 
In the northern regions, efforts should focus on mitigating alcohol-related risks by implementing awareness campaigns, strengthening preventive education in schools and workplaces, or enforcing stricter regulations on alcohol availability and marketing.

\section{Conclusion and future research directions}
\label{discussion}

This study proposes a comprehensive statistical framework to examine how occupational factors in combination with socioeconomic conditions influence health-related risky behaviors in Italy. By integrating structural topic modeling with dynamic multivariate ordinal probit models, it allows for a detailed characterization of behavioral heterogeneity and its temporal evolution. The complexity of the proposed modeling strategy mirrors the intrinsic complexity of the phenomenon under investigation and aligns with a growing movement in epidemiology that emphasizes the importance of embracing complexity \citep{pearce2006complexity, 
stronks2018embracing, rod2023complexity}.
In the existing literature, analyses typically focus either on socioeconomic determinants or on occupational characteristics, often treating them as separate domains. In contrast, our approach reveals that such relationships are strongly modulated by occupational conditions and evolve over time, underscoring the importance of modeling contextual and temporal heterogeneity explicitly.
It provides individualized analyses based on representative respondents and can inform ongoing surveillance of behavioral risk factors, allowing analyses during periods of limited data availability (e.g., the COVID-19 pandemic) and the identification of trend changes (e.g., before and after the COVID-19 pandemic).

Despite its strengths, several limitations remain. 
Macro-territoriality has been included within the model through dummy components, ensuring that broad spatial differences are accounted for. 
However, the surveillance system could also support finer-grained analyses at the micro level through spatial models. Achieving this would require further methodological effort in model parameterization and the development of dedicated algorithms for spatio-temporal inference. The current setup performs filtering rather than full and retrospective smoothing, meaning that continuous updating of estimates is possible as new data arrive, but additional methodological efforts are needed for backward dynamic inference. Although the model is designed at the individual level, policy interest may also lie in macro and aggregated groups. 
Since the model operates on individuals, upward aggregation is possible, allowing users or policymakers to dynamically create higher-level groupings when needed.
However, such operations entail substantial computational effort.
To make the model’s results fully accessible and actionable, a simplified and user-friendly analytical interface would be required \citep{stival2025communicatingcomplexstatisticalmodels}. These developments, together with further optimization of computational routines, are left for future research.

In conclusion, while further extensions are warranted, the proposed approach addresses a major gap in the joint analysis of health behaviors and occupational contexts in Italy. Moreover, the methodology offers a general template for other countries with comparable behavioral risk-factor surveillance systems.

\section*{Acknowledgments}
The authors are grateful to the So.Sta. group (A. Arletti, M. Bertani, G. Bertarelli, M. Marzulli, A. Pastore, M. Pittavino, S. Tonellato) at Ca’ Foscari University of Venice for their insightful comments. We also thank the Gruppo Tecnico PASSI (PASSI national coordination team at ISS), and the PASSI network interviewers.

\bibliographystyle{apalike} 
\bibliography{bibliography}       

\newpage


\begin{appendix}

\makeatletter
\renewcommand\thefigure
{\@arabic\c@figure}
\def\fps@figure{tbp}
\def\ftype@figure{2}
\def\fstyle@figure{\reset@font\small\rm}
\def\ext@figure{lof}
\def\fnum@figure{{\bf Figure S\thefigure}
}
\makeatother

\makeatletter
\def\fnum@table{\textbf{Table S\thetable}}
\makeatother

\renewcommand{\thesection}{Appendix \Alph{section}}

\section*{\huge Supplementary Material}
\ref{appA} includes job-related covariate information. \ref{appB} reports details on spike-and-slab dynamic prior specified on regression coefficients.
In \ref{appC} computational aspects of the SMC algorithm are reported.
\ref{appD} provides additional details and results of the simulation experiments. \ref{appE} reports computational and prior settings and additional results related to the Italian PASSI data application. 

\section{Job variables}\label{appA}

Table~\ref{tab:sector_labels} shows the classification of occupational sectors, while Table~\ref{tab:ISTAT_labels} refers to the job classification according to the ISTAT (\url{https://professioni.istat.it/sistemainformativoprofessioni/cp2011/}).

\begin{table}[ht]
\centering
\caption{Occupational sector classification labels}\label{tab:sector_labels}
\begin{tabular}{rl}
\toprule
\textbf{Value} & \textbf{Label} \\
\midrule
1  & Agriculture \\
2  & Food industry \\
3  & Metalworking industry \\
4  & Electrical-electronic industry \\
5  & Textile and clothing industry \\
6  & Chemical and ceramic industry \\
7  & Wood and paper industry \\
8  & Other manufacturing industries \\
9  & Construction \\
10 & Energy, gas, water, telecommunications \\
11 & Trade and public services \\
12 & Transport \\
13 & Banking and insurance \\
14 & School and university \\
15 & Health care \\
16 & Public administration \\
17 & Business services \\
18 & Personal services \\
19 & Law enforcement/military \\
\bottomrule
\end{tabular}
\end{table}

\begin{table}[ht]
\centering
\caption{ISTAT job classification labels}
\label{tab:ISTAT_labels}
\begin{tabular}{rl}
\toprule
\textbf{Value} & \textbf{Label} \\
\midrule
1  & Entrepreneurs, senior executives, and legislators \\
2  & Intellectual, scientific, and highly specialized professions \\
3  & Technical professions \\
4  & Clerical support workers \\
5  & Skilled workers in commerce and services \\
6  & Craftspeople, skilled workers, and farmers \\
7  & Plant operators, machinery workers, and drivers \\
8  & Elementary occupations \\
9  & Armed forces \\
\bottomrule
\end{tabular}
\end{table}

\begin{figure}[ht]
   \centering
    \includegraphics[width=.7\linewidth]{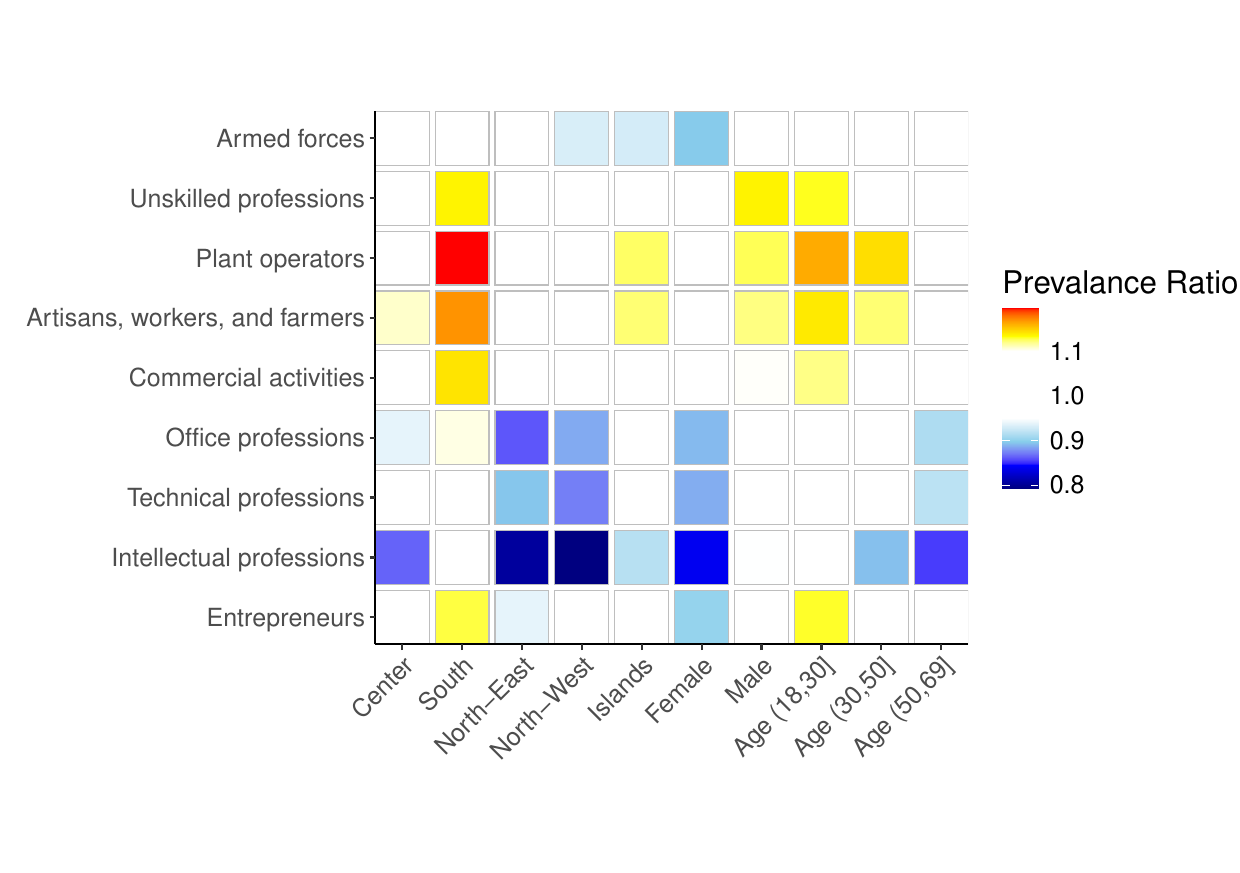}
    \caption{Prevalence ratio between the group-specific prevalence and the overall population prevalence of exhibiting at least one risky behavior by demographic variable ($x$-axis) and ISTAT job category ($y$-axis).
    }
    \label{fig:heatmap}
\end{figure}

Figure~\ref{fig:heatmap} illustrates how the observed ratio between the prevalence of individuals with at least one risky behavior within a given group and the corresponding prevalence in the total population varies across demographic groups and job categories, the latter defined by \cite{istat_cp2011}, i.e., Table \ref{tab:ISTAT_labels}. Distinct patterns emerge: intellectual workers tend to adopt healthier lifestyles, whereas manual laborers show a higher prevalence of risky behaviors, especially in southern regions.

\newpage

\section{Details on spike-and-slab dynamic prior}\label{appB}

According to \citet{johnson2010}, who first introduced nonlocal priors, 
a density $f_{\textnormal{slab}}(\beta)$ is nonlocal on a subspace 
 $\mathcal{S}_\beta^0$ of the support $\mathcal{S}_\beta$ if, 
 for every $\epsilon > 0$, there exists a neighborhood around $\beta_0$ 
 within $\mathcal{S}_\beta^0$ such that $f_{\textnormal{slab}}(\beta) < \epsilon$.
 In our setting, $\mathcal{S}_\beta = \mathbbm{R}$ and $\beta_0 = 0$, 
 implying that the nonlocal slab exhibits two modes away from zero. 

Considering the specification of the slab $f_\textnormal{slab}(\beta_{kpq}^{(t)})$ 
 as a bimodal mixture of two Gaussians $\mathcal{N}(\mu_{-1}, \sigma_{-1})$ and $\mathcal{N}(\mu_{1}, \sigma_{1})$, multiplied by a suitable weighted function $\omega(\beta_{kpq}^{(t)}; \xi)$ that reduces mass near the origin.
 We let $\rho(\beta_{kpq}^{(t-1)}): \mathbb{R}\rightarrow (0,1)$ denote the mixing probability and we set   
    $$
    \omega(\beta_{kpq}^{(t)}; \xi)=1-e^{-(\beta_{kpq}^{(t)}/\xi)^2}.
    $$

Let $\boldsymbol{\mu} = (\mu_{-1}, \mu_{1})^\top$ and $\boldsymbol{\sigma} = (\sigma_{-1}, \sigma_{1})^\top$ denote the mean and standard deviation vectors of the two Gaussian components.
The resulting nonlocal slab density is
\begin{align*}
    f_{\textnormal{slab}}(\beta_{kpq}^{(t)};\boldsymbol{\mu}, \boldsymbol{\sigma}, \beta_{kpq}^{(t-1)}, \xi) =  \frac{\omega(\beta_{kpq}^{(t)}; \xi)}{c(\boldsymbol{\mu}, \boldsymbol{\sigma}, \beta_{kpq}^{(t-1)}, \xi)} [\{1-&\rho(\beta_{kpq}^{(t-1)}) \} f_N(\beta_{kpq}^{(t)}; \mu_{-1}, \sigma_{-1}^2)+\\
    &\rho(\beta_{kpq}^{(t-1)}) f_N(\beta_{kpq}^{(t)}; \mu_{1}. \sigma_{1}^2)],
\end{align*}
where $c(\boldsymbol{\mu}, \boldsymbol{\sigma}, \beta_{kpq}^{(t-1)}, \xi)$ is the normalizing constant.

Then we obtain
\begin{align*}
p(\beta_{kpq}^{(t)} \vert \beta_{kpq}^{(t-1)}) &= 
\pi_{0}(\beta_{kpq}^{(t-1)})  f_{N}(\beta_{kpq}^{(t)}; 0, \sigma_0^2) + \sum_{l \in \{-1,1\}} \pi_{l}(\beta_{kpq}^{(t-1)})
     \frac{\omega(\beta_{kpq}^{(t)}; \xi) f_N(\beta_{kpq}^{(t)}; \mu_l, \sigma_l^2)}     {c(\boldsymbol{\mu}, \boldsymbol{\sigma}, \rho, \xi)},
\end{align*}
with kernel functions
\begin{align*}
g_{-1}(\beta_{kpq}^{(t)}; \mu_{-1}, \sigma_{-1}, \xi)&= \frac{\omega(\beta_{kpq}^{(t)}; \xi) f_N(\beta_{kpq}^{(t)}; \mu_{-1}, \sigma_{-1}^2)}     {c(\boldsymbol{\mu}, \boldsymbol{\sigma}, \rho, \xi)},\\
g_{1}(\beta_{kpq}^{(t)}; \mu_{1}, \sigma_{1}, \xi) &=  \frac{\omega(\beta_{kpq}^{(t)}; \xi) f_N(\beta_{kpq}^{(t)}; \mu_1, \sigma_1^2)}     {c(\boldsymbol{\mu}, \boldsymbol{\sigma}, \rho, \xi)},
\end{align*}
and mixture
weights
\begin{align*}
    \pi_{-1}(\beta_{kpq}^{(t-1)}) &= \{1- \pi_0(\beta_{kpq}^{(t-1)})\}\{1-\rho(\beta_{kpq}^{(t-1)})\},\\
    \pi_{1}(\beta_{kpq}^{(t-1)}) &= \{1- \pi_0(\beta_{kpq}^{(t-1)})\}\rho(\beta_{kpq}^{(t-1)}), \quad \sum_{l=-1}^1 \pi_l(\beta_{kpq}^{(t-1)}) = 1.
\end{align*}

\begin{figure}[ht]
    \centering
    \includegraphics[width=.8\linewidth]{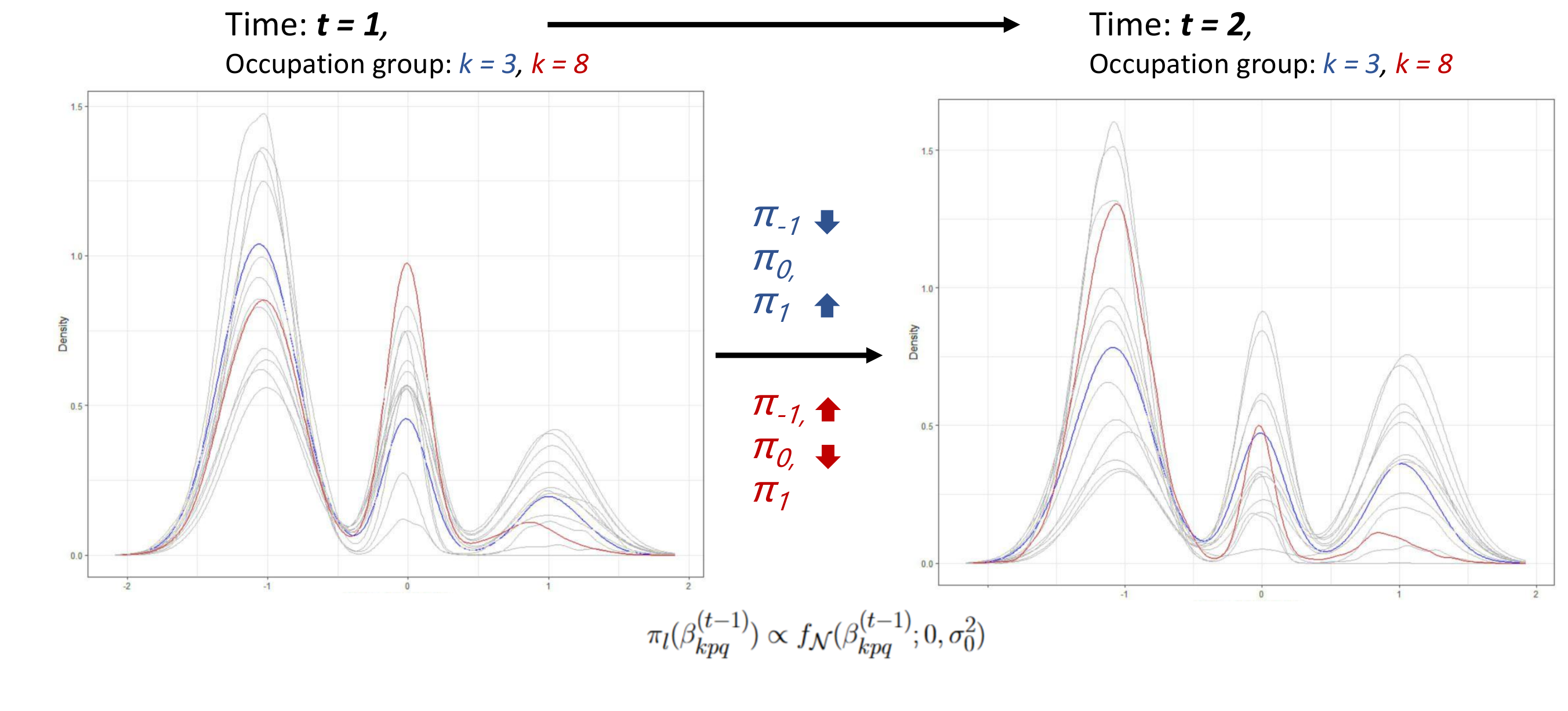}
    \caption{Example of prior density evolution for regression coefficients $\beta_{kpq}^{(1)}$--$\beta_{kpq}^{(2)}$ for two groups $k\in\{3,8\}$.}
    \label{fig:beta-evo}
\end{figure}

To model coefficient dynamics, we specify the prior mixture probabilities $\pi_{l}(\beta_{kpq}^{(t-1)})$ at time $t>1$ as
\begin{align*}
    \pi_{0}(\beta_{kpq}^{(t-1)}) &= f_\mathcal{N}(\beta_{kpq}^{(t-1)}; 0, \sigma_0^2)/\, \varpi, \quad  \pi_{l}(\beta_{kpq}^{(t-1)}) = g_{l}(\beta_{kpq}^{(t-1)}; \mu_{l}, \sigma_{l}, \xi)/\, \varpi, \quad 
\end{align*}
for $l \in \{-1,1\}$, with  $\varpi = f_\mathcal{N}(\beta_{kpq}^{(t-1)}; 0, \sigma_0^2) + \sum_{l \in \{-1,1\}}   g_{l}(\beta_{kpq}^{(t)}; \mu_{l}, \sigma_{l}, \xi)$. If $t=1$, we set fixed hyperparameters $\pi_{l}(\beta_{kpq}^{(1)})$, for $l \in \{-1,0,1\}$.
Figure \ref{fig:beta-evo} shows simple examples of dynamics in prior density of the entries of the coefficient matrix $\boldsymbol{B}$ as defined above.

\newpage

\section{Computational details of SMC algorithms}
\label{appC}

In  Algorithm 2, the Metropolis-Hastings acceptance ratio $\alpha$ is computed in Step~IIIii by considering the posterior density of $\boldsymbol{B}^{(t)}$  given $\mathcal{D}^{(1:t-1)}$ and the first $n$ observations in period $t$, with $\boldsymbol{s}^{(1:t)}$ marginalized out.  
This quantity can be expressed as  
\begin{align*}
    p(\boldsymbol{B}^{(t)}_{j} \vert \boldsymbol{Y}^{(1:t-1)}, \boldsymbol{y}^{(t)}_{1:n}
    \boldsymbol{X}^{(1:t)}, \boldsymbol{W}^{(1:t)}) \propto
    &p(\boldsymbol{B}^{(t)}_{j} \vert \mathcal{D}^{(1:t-1)} ) \prod_{m=1}^{n} p(\boldsymbol{y}_{m} \vert \boldsymbol{B}^{(t)}_{j} , \boldsymbol{X}^{(t)}, \boldsymbol{W}^{(t)}, \boldsymbol{V}^{(t)} ),
\end{align*}
where the likelihood term $p(\boldsymbol{y}_{m} \vert \boldsymbol{B}^{(t)}_{j} , \boldsymbol{X}^{(t)}, \boldsymbol{W}^{(t)}, \boldsymbol{V}^{(t)} )$  
is available in closed form as a mixture over the groups $k=1,\ldots, K$.  
Notably, the product of likelihood terms corresponds to the joint distribution of $(\boldsymbol{y}^{(t)}_{1}, \ldots, \boldsymbol{y}^{(t)}_{n})$,  
where independence follows from the assumptions in Subsection 3.1.

New values of $\tilde{\boldsymbol{B}}^{(t)}_{j}$ (proposal) are obtained by either considering the prior distribution of $\boldsymbol{B}^{(1)}$ or by jittering the current value.  
The prior density $p(\boldsymbol{B}^{(t)}_{j} \vert \mathcal{D}^{(1:t-1)})$ is evaluated at the current state $\boldsymbol{B}^{(t)}_{j}$ using results from the previous iteration $n-1$, whereas its value at the proposal $\tilde{\boldsymbol{B}}^{(t)}_{j}$ is computed via numerical integration. 
Specifically, we approximate the integral  
\begin{equation*}
    p(\tilde{\boldsymbol{B}}^{(t)}_{j} \vert \mathcal{D}^{(1:t-1)} ) = \int_B p(\tilde{\boldsymbol{B}}^{(t)}_{j} \vert \boldsymbol{B}^{(t-1)} ) \, \textnormal{d}  p( \boldsymbol{B}^{(t-1)}_{j} \vert \mathcal{D}^{(1:t-1)} )
\end{equation*}
using the Monte Carlo average of $p(\tilde{\boldsymbol{B}}^{(t)}_{j} \vert \boldsymbol{B}^{(t-1)}_{h} )$ over a set of $H$ particles $\boldsymbol{B}^{(t-1)}_{h}$, where $h=1,\ldots, H$. These particles are sampled from the filtering distribution $p(\boldsymbol{B}^{(t-1)} \vert \mathcal{D}^{(t-1)})$ at the previous iteration $t-1$ of Algorithm 1.

At each time period $t$, one could instead rely on MCMC-based posterior approximation schemes.
In \cite{goudie2019joining}, samples are obtained sequentially by considering 
$$ p(\boldsymbol{s}^{(t)}, \boldsymbol{B}^{(t)} \vert \mathcal{D}^{(1:t)}) \propto p(\boldsymbol{s}^{(t)}, \boldsymbol{\theta}^{(t)}, \boldsymbol{B}^{(t)} \vert \mathcal{D}^{(t)}, \boldsymbol{B}^{(t-1)}) p(\boldsymbol{B}^{(t-1)} \vert \mathcal{D}^{(1:t-1)}), $$
with $p(\boldsymbol{B}^{(t-1)} \vert \mathcal{D}^{(1:t-1)})$ given in particles.

\section{Details on simulation experiments and additional results}\label{appD}

To set  $\boldsymbol{X}$, $\bar{\boldsymbol{B}}^{(1)}$ in each scenario of the simulation experiments, we proceed as follows: for each period $t$, we define the binary covariate matrices $\boldsymbol{X}^{(t)}$ with dimensions $N_t \times Q$, where the first column is set to one---allowing for the intercept coefficients---and the remaining  $Q - 1$ columns sampled independently from a Bernoulli distribution with success probability $0.5$.
The entries of the first-period coefficient matrix $\bar{\boldsymbol{B}}^{(1)}$ are initially sampled from a Gaussian distribution $\mathcal{N}(0, 2)$.
To enforce exact sparsity, a thresholding step is then applied, setting entries with absolute values below the empirical quantile of level $\varsigma$ to zero.
Temporal dynamics across periods are introduced by sampling the entries of $\bar{\boldsymbol{B}}^{(2)}$ from $\mathcal{N}(\bar{\beta}_{kpq}^{(1)}, 0.5)$ and, again, thresholding them to zero according to the $\varsigma$-level empirical quantile of the drawn entries in absolute value. 
Group memberships $\bar{s}_n$ for the synthetic data are drawn from a categorical distribution with probabilities $\bar{\boldsymbol{\theta}}_n$, where $\bar{\boldsymbol{\theta}}_n$ is sampled from a Dirichlet distribution with parameter vector $(\exp(\bar{\boldsymbol{\eta}}_n), 1)^\top$. 
To emulate the inference task in the motivating application, where posterior inference on $\boldsymbol{\theta}_n$ is first conducted via topic modeling, all competing algorithms in these experiments are provided with the (true) values of $\bar{\boldsymbol{\eta}}_n$ and $\Sigma_\eta$, treated as a given.

After setting $\boldsymbol{X}$, $\bar{\boldsymbol{B}}^{(1)}$, $\bar{\boldsymbol{B}}^{(2)}$, and $\bar{\boldsymbol{s}}$, we replicate $R = 20$ times the generation of response variables $\boldsymbol{Y}^{(1)}$ and $\boldsymbol{Y}^{(2)}$ according to the ordinal multivariate mixture model in Equation (1). For $y_p$ with $p \in \{1,2\}$, we set $C_p = 3$ and $\boldsymbol{\tau}_p = (-0.5, 0.5)^\top$, while for $p \in \{3,4\}$, we set $C_p = 4$ and $\boldsymbol{\tau}_p = (-0.75, 0, 0.75)^\top$.

We specify the following prior on $\boldsymbol{B}^{(1)},\boldsymbol{B}^{(2)}$.
Each entry $\beta_{jkq}^{(t)}$ is distributed according to nonlocal spike-and-slab prior defined in Equation 5, with $(\mu_1, \mu_{-1})=(-1.5,1.5)$, $(\sigma_{1}, \sigma_0, \sigma_{-1})=(0.75,0.1,0.75)$, and $\xi=2$.

\begin{figure}[ht]
    \centering
    \includegraphics[width=.5\linewidth]{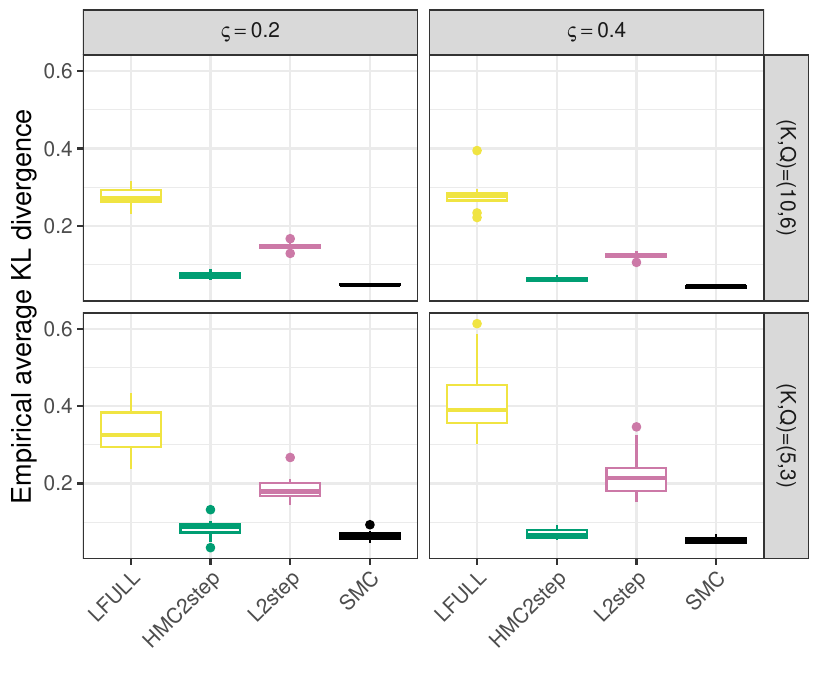}
    \caption{Boxplots of the distribution over $R = 20$ simulation replications of the average KL divergence between the benchmark predictive posterior samples provided by \textsc{HMCFull} and those generated by the competing models, evaluated across varying sparsity levels $\varsigma$ and parameter dimensions $K$ and $Q$. The number of subjects is $N_1=250$ and $N_2 = 300$.}
    \label{fig:sim-predSM}
\end{figure}

\begin{figure}[ht]
    \centering
    \includegraphics[width=.5\linewidth]{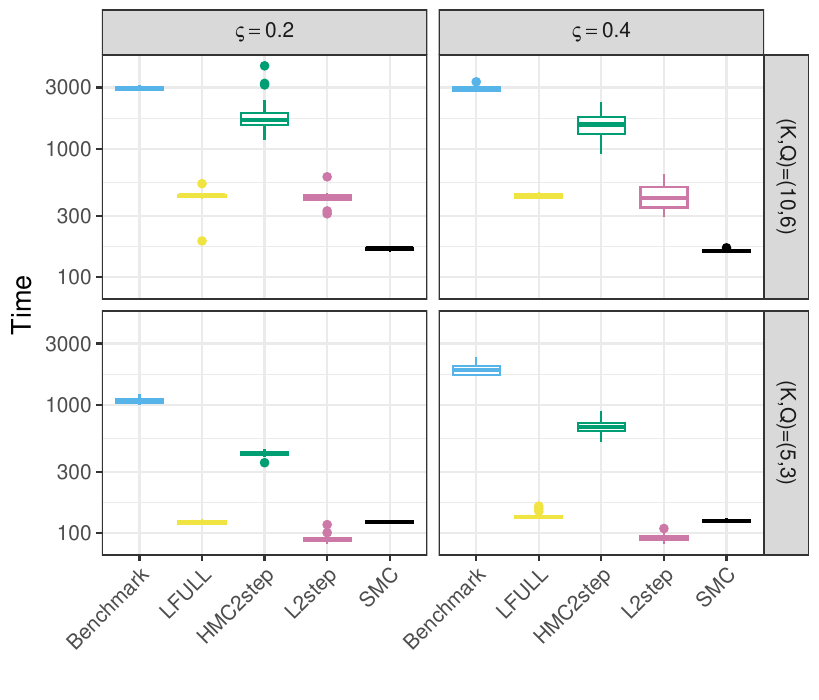}
    \caption{Boxplots of the distribution across $R = 20$ simulation replications of the time (in seconds with axis in base-10 logarithmic scale) required to generate predictive posterior samples for subjects at $t = 2$ for each competing model, evaluated across varying sparsity levels $\varsigma$ and parameter dimensions $K$ and $Q$. The benchmark indicates the log-time distribution associated with \textsc{HMCFull}. The number of subjects is $N_1=250$ and $N_2 = 300$. }
    \label{fig:sim-timeSM}
\end{figure}

\newpage

\section{Additional application settings and results}
\label{appE}

We train the STM using data collected up to December~2019, i.e.,  $t^{\star} = 72$,
and set $(\boldsymbol{\Lambda} = \hat{\boldsymbol{\Lambda}}_{t^{\star}}$, $\boldsymbol{\Psi})=\hat{\boldsymbol{\Psi}}_{t^{\star}}$, and 
$\boldsymbol{\Gamma})=\hat{\boldsymbol{\Gamma}}_{t^{\star}}$. 
Prior elicitation is completed by setting $(\mu_1. \mu_{-1})=(-1.5, 1.5)$, $(\sigma_{1},\sigma_{0},\sigma_{-1})=(0.75,0.10, 0.75)$, and $\xi=2$.

We employ parallel computing by running 20 parallel instances with 150 particles each, yielding a total of 
$J = 3000$ particles to approximate the posterior predictive distribution at 
each month $t$. Filtering a single observation requires approximately $3.5$ seconds on a Windows machine equipped with an AMD EPYC~7413~24-Core Processor (2.65~GHz), using 20 cores.

Table \ref{tab:clusters} shows the representative subjects identified by the \texttt{findThoughts} function from the \texttt{stm} package~\citep{stm-r} for each of the 30 clusters obtained from the topic model. For each cluster, the table reports the representative text $\mathbf{w}_{\tilde{n}}$, the corresponding socio-economic covariates $\mathbf{v}_{\tilde{n}}$ (i.e., economic problems and educational level), and an interpretative label summarizing these three pieces of information.

\setlength{\tabcolsep}{4pt}
\renewcommand{\arraystretch}{1.15}

\begin{center}
\footnotesize
\begin{longtable}{r>{\raggedright\arraybackslash}p{5cm}%
                  >{\centering\arraybackslash}p{1.5cm}%
                  >{\centering\arraybackslash}p{1cm}%
                  >{\raggedright\arraybackslash}p{6cm}}
\caption{Description of the representative subjects founded by the \texttt{findThoughts} function from the \texttt{stm} package. The label is made considering the related text and socio-economic covariates used in the STM.}\label{tab:clusters}\\
\toprule
\textbf{Cluster} & \textbf{Text} & \textbf{Economic Problems} & \textbf{Edu. Level} & \textbf{Label}\\
\midrule
\endfirsthead

\toprule
\multicolumn{5}{l}{\small\textit{(continued)}}\\
\textbf{Cluster} & \textbf{Text} & \textbf{Economic Problems} & \textbf{Edu. Level} & \textbf{Label}\\
\midrule
\endhead

\midrule
\multicolumn{5}{r}{\small\textit{(continues next page)}}\\
\endfoot

\bottomrule
\endlastfoot

1  & infermier intens ospedal professional san tecnic terap & 1 & High & Nurse, hospital care, higher educated, with economic problems\\
2  & coll domest person qualific serviz tric & 1 & Low & Domestic helper/cleaner, lower educated, with economic problems\\
3  & agricoltor agricoltur artigian coltiv dirett opera specializz & 1 & Low & Farmer and agricultural worker, lower educated, with economic problems\\
4  & serviz addett attes attiv commercial fuoc person qualific soccors tip vigil & 0 & High & Security and emergency services worker, higher educated, without economic problems\\
5  & can commerc eserciz pubblic qualific toilett & 1 & Low & Dog groomer, lower educated, with economic problems\\
6  & aci adegu attual autist conducent conduttor fa impiant macchinar opera precar trasport trov veicol & 1 & High & Driver and vehicle operator, higher educated, with economic problems\\
7  & san tecnic & 1 & Low & Healthcare worker, lower educated, with economic problems\\
8  & forz armat costier gurd imbarc militar mot ordin vedett & 0 & High & Military/coast guard, higher educated, without economic problems\\
9  & serviz attiv commercial ipercoop person qualific & 1 & Low & Retail sales worker, lower educated, with economic problems\\
10 & aliment bei chauffeur der industr qualific senn & 0 & Low & Food industry worker, lower educated, without economic problems\\
11 & asil assintent nid qualific scuol univers & 1 & Low & Early childhood assistant, lower educated, with economic problems\\
12 & cart legn mobil montagg qualific & 0 & High & Forestry and woodworking craftsperson, higher educated, without economic problems\\
13 & assicur banc brokeragg esecut impieg lavor uffic & 0 & High & Insurance and banking clerk, higher educated, without economic problems\\
14 & esecut lavor uffic amministr d professioon pubblic & 0 & High & Public administration clerk, higher educated, without economic problems\\
15 & industr metalmeccan operaiain qualific & 0 & High & Skilled metal industry worker, higher educated, without economic problems\\
16 & person serviz comun qualific tram & 1 & High & Worker in community personal services, higher educated, with economic problems\\
17 & dirigent agricol agricoltur alta azienz imprenditor legisl ris & 0 & High & Agricultural entrepreneur, higher educated, without economic problems\\
18 & tessil abbigl idustr operai qualific & 1 & High & Skilled textile industry worker, higher educated, with economic problems\\
19 & x azi consulent cont ditt esecut finanziar impieg impres lavor priv pubblic revision serviz uffic & 1 & High & Financial and administrative office clerk, higher educated, with economic problems\\
20 & cant ediliz geometr tecnic & 0 & High & Construction site technician, higher educated, without economic problems\\
21 & acqua call center energ gas tecnic telefon & 1 & High & Utilities call-center, higher educated, with economic problems\\
22 & attrezz ceram chimic stamp tecnic & 1 & High & Employee in the ceramics and chemical sector, higher educated, with economic problems\\
23 & anzian assistent assunt person puliz qualific serviz & 1 & Low & Elderly care and cleaning staff, lower educated, with economic problems\\
24 & elettr elettric elettron industr plc programm tecnic & 0 & High & Industrial electrician/technician, higher educated, without economic problems\\
25 & elev intellettual scientif specializz profession scuol univers & 0 & High & Qualified educator, higher educated, without economic problems\\
26 & ediliz mont qualific taparell zanzar & 1 & Low & Installation worker in construction, lower educated, with economic problems\\
27 & attiv camer commerc commercial eserciz fast food pubblic qualific serviz & 0 & Low & Fast food server, lower educated, without economic problems\\
28 & agost and april attiv bed breakfast camerier commerc commercial dicembr eserciz lugl pubblic qualific serviz stagional & 1 & High & Seasonal waitress, higher educated, with economic problems\\
29 & tecnic atrezzatur cosulent lavor person serviz sicurezz & 0 & High & Occupational safety technician, higher educated, without economic problems\\
30 & amministr manutenzion pubblic qualific spaz verd & 1 & High & Green area maintenance worker, higher educated, with economic problems\\
\end{longtable}
\end{center}

\begin{figure}[ht]
    \centering
    \includegraphics[width=.5\linewidth]{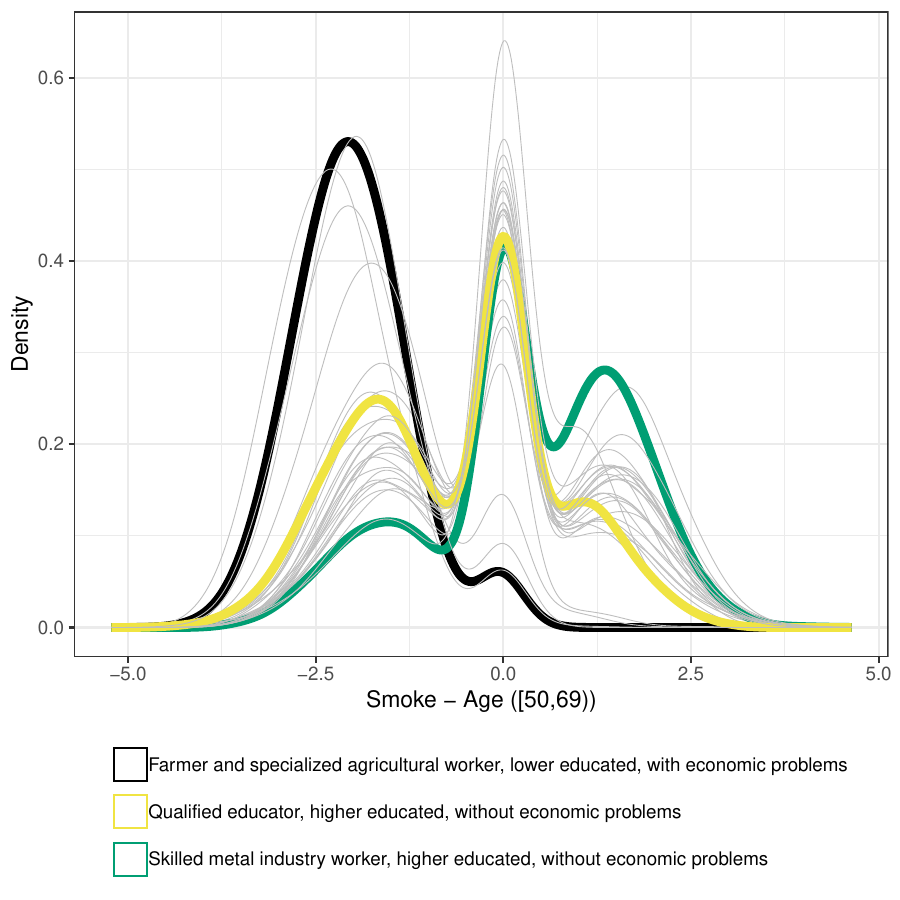}
    \caption{Posterior distributions of the coefficients $\beta^{(1)}_{1k3}$ ($k = 1, \ldots, 30$) associated with the $(50,69]$ age class and smoking-related risky behavior.}
    \label{fig:density}
\end{figure}

Figure~\ref{fig:density} displays the posterior distributions of the coefficients $(\beta^{(1)}_{1k3}$, $k = 1, \ldots, K$), revealing substantial heterogeneity across latent groups in the impact of belonging to the $(50,69]$ age class on smoking-related behavior.
At $t = 1$, older workers are more likely to smoke in skilled metal industries, while the reverse holds for agricultural occupations.

\begin{figure}[ht]
    \centering
    \includegraphics[width=\linewidth]{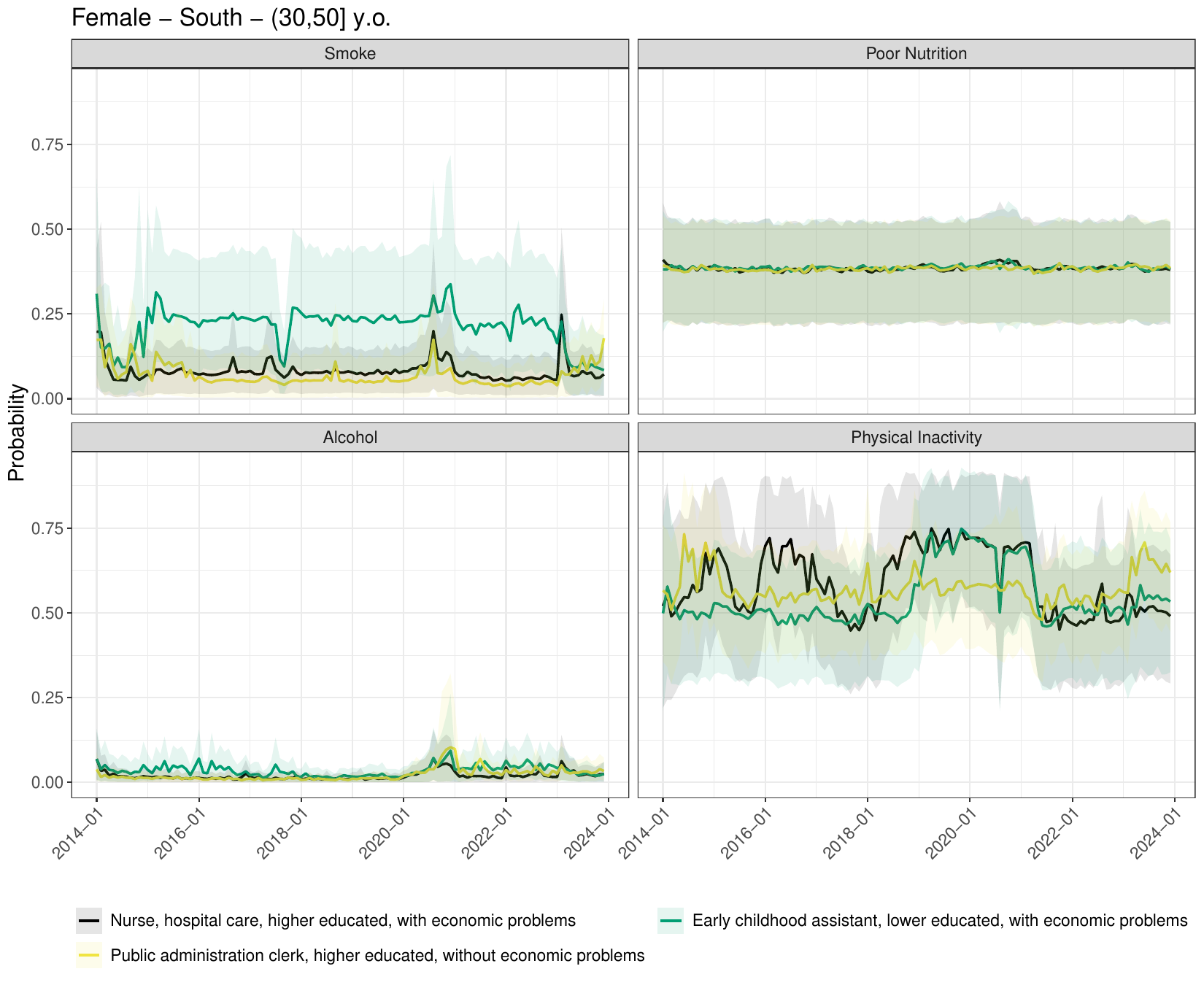}
    \caption{Estimated risk probabilities by year for each SNAP behavior, for a hypothetical 30–50-year-old female from south Italy and three latent job groups.}
    \label{fig:ts1}
\end{figure}

Figure~\ref{fig:ts1} shows the temporal trends of smoking risk probabilities for a female aged 30–50 from southern Italy.
Individuals working in early childhood education and care settings display a higher propensity to smoke throughout the observed period, although in the most recent years their risk converges to the levels observed in the other two groups, characterized by higher education and, in one case, better economic status. All three job groups exhibit a rising propensity for physical inactivity during the COVID-19 pandemic.

Figure~\ref{fig:heatmap_alcohol} shows the relative risk associated with alcohol consumption, following Figure~5 from the main manuscript.
We can observe that being from the North-East increases this risk, particularly for qualified educators, agricultural entrepreneurs, and industrial electricians, characterised by high educational and economic levels.
Conversely, being from the South appears to have a mainly protective effect, remembering that the baseline category corresponds to individuals from Central Italy.

\begin{figure}[ht]
    \centering
    \includegraphics[width=.8\linewidth]{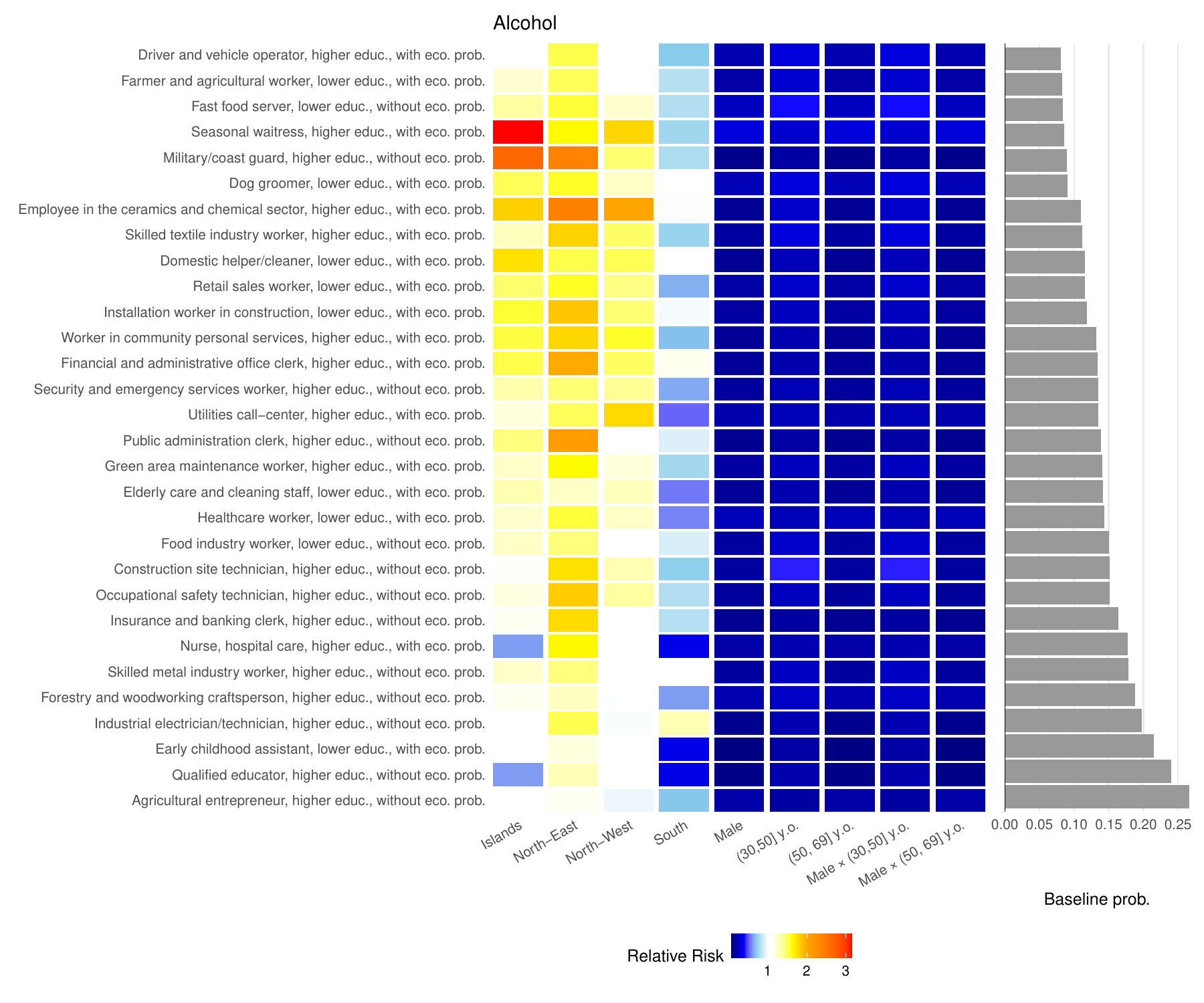}
    \caption{Alcohol relative risk for each job group and socio-demographic covariates for December 2023.}
    \label{fig:heatmap_alcohol}
\end{figure}

\begin{figure}[ht]
    \centering
    \includegraphics[width=.9\linewidth]{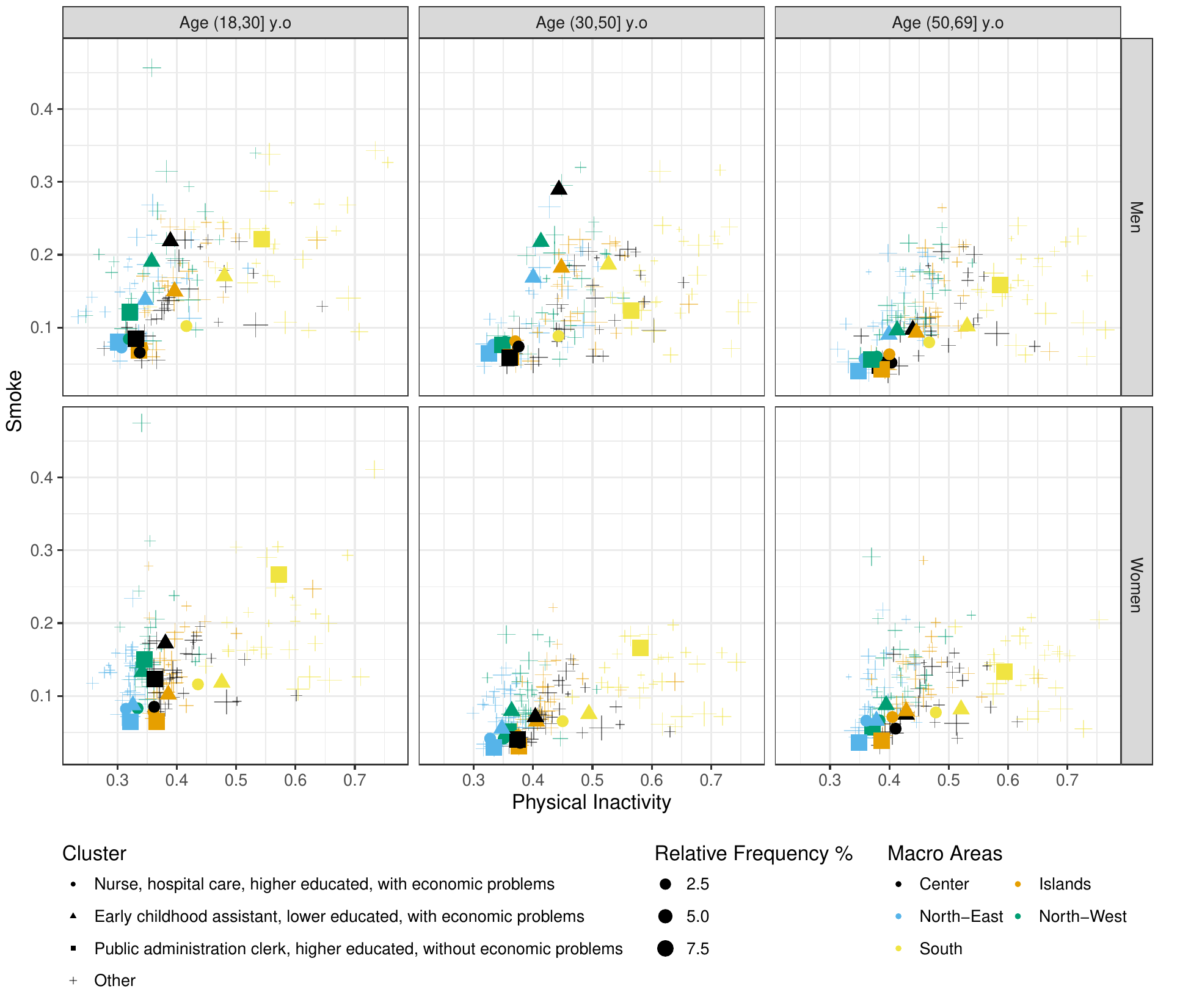}
    \caption{Estimated probabilities of engaging in risky smoke use and physical inactivity, by sex, age group, macro area, and three latent job groups.}
    \label{fig:pm2}
\end{figure}

Finally, Figure~\ref{fig:pm2} displays the estimated probabilities of smoking and physical inactivity across socio-demographic variables and job classes.
Individuals from the South working in public administration show high levels of risk for both behaviors, particularly among younger age groups.
Overall, being from the South (colored in yellow) appears to act as a risk factor for unhealthy habits such as smoking and physical inactivity across different job groups.

\end{appendix}

\end{document}